\documentclass[numbers]{sigplanconf}

\usepackage{amsmath}
\usepackage{graphicx, subfig}
\usepackage{caption}
\usepackage{booktabs}
\usepackage{placeins}
\usepackage{multirow}
\usepackage{url}
\usepackage{microtype}
\usepackage{algorithmicx,algorithm,algpseudocode}
\usepackage[export]{adjustbox}
\usepackage{enumitem}
\usepackage{pdfpages}
\usepackage{verbatim}

\usepackage{ifthen,color}
\definecolor{darkgreen}{rgb} {0.0,0.5,0.0}
\definecolor{darkblue}{rgb} {0.0,0.0,0.5}
\definecolor{bluegreen}{rgb}{0.0,0.5,0.5}

\newcommand{\yzh}[1]{{\it{\color{darkblue}(YZH) #1}}}
\newcommand{\john}   [1]{{{\color{darkgreen}(John) #1}}}

\newcommand{\final}{0}
\ifthenelse{\equal{\final}{1}}{\renewcommand{\john}[1]{}\renewcommand{\yzh}[1]{}}{}
\newcommand{\includehelptoggle}{0}
\ifthenelse{\equal{\includehelptoggle}{1}}{\newcommand{\includehelp}[1]{#1}}{\newcommand{\includehelp}[1]{}}

\ifdefined\textln\relax\else\newcommand{\textln}[1]{#1}\fi

\usepackage{listings}
\usepackage{color}

\definecolor{dkgreen}{rgb}{0,0.6,0}
\definecolor{gray}{rgb}{0.5,0.5,0.5}
\definecolor{mauve}{rgb}{0.58,0,0.82}

\usepackage[firstpage]{draftwatermark}
\SetWatermarkText{\hspace*{8in}\raisebox{6.5in}{\includegraphics{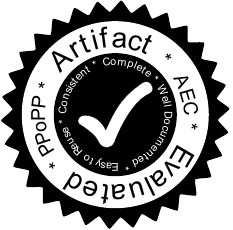}}}
\SetWatermarkAngle{0}

\begin{document}

\setlength{\pdfpageheight}{\paperheight}
\setlength{\pdfpagewidth}{\paperwidth}

\copyrightyear{2016}
\conferenceinfo{PPoPP '16}{March 12-16, 2016, Barcelona, Spain}
\copyrightdata{978-1-4503-4092-2/16/03}
\reprintprice{\$15.00}
\copyrightdoi{2851141.2851145}
\publicationrights{licensed}





\title{Gunrock: A High-Performance Graph Processing Library on the GPU}
\subtitle{}

\authorinfo{Yangzihao Wang, Andrew Davidson\titlenote{Currently an employee at Google.}, Yuechao Pan, Yuduo Wu\titlenote{Currently an employee at IBM\@.}, Andy Riffel, John D. Owens}
      {University of California, Davis}
     {\{yzhwang, aaldavidson, ychpan,  yudwu, atriffel, jowens\}@ucdavis.edu}

\maketitle

\begin{abstract}
For large-scale graph analytics on the GPU, the irregularity of data
access/control flow and the complexity of programming GPUs have been two
significant challenges for developing a programmable high-performance graph
library. ``Gunrock,'' our high-level bulk-synchronous graph-processing system
targeting the GPU, takes a new approach to abstracting GPU graph analytics:
rather than designing an abstraction around \emph{computation}, Gunrock instead
implements a novel \emph{data-centric} abstraction centered on operations on
a vertex or edge frontier. Gunrock achieves a balance between performance and
expressiveness by coupling high-performance GPU computing primitives and
optimization strategies with a high-level programming model that allows
programmers to quickly develop new graph primitives with small code size and
minimal GPU programming knowledge.  We evaluate Gunrock on five graph
primitives (BFS, BC, SSSP, CC, and PageRank) and show that Gunrock has on
average at least an order of magnitude speedup over Boost and PowerGraph,
comparable performance to the fastest GPU hardwired primitives, and better
performance than any other GPU high-level graph library.
\end{abstract}




\section{Introduction}
\label{sec:intro}

Graphs are ubiquitous data structures that can represent relationships between
people (social networks), computers (the Internet), biological and genetic
interactions, and elements in unstructured meshes, just to name a few. In this
paper, we describe ``Gunrock,'' our graphics processor (GPU)-based system for
graph processing that delivers high performance in computing graph analytics
with its high-level, data-centric parallel programming model. Unlike previous
GPU graph programming models that focus on sequencing computation steps, our
data-centric model's key abstraction is the \emph{frontier}, a subset of the
edges or vertices within the graph that is currently of interest. All Gunrock
operations are bulk-synchronous and manipulate this frontier, either by
computing on values within it or by computing a new frontier from it.

At a high level, Gunrock targets graph primitives that are iterative,
convergent processes. Among the graph primitives we have implemented
and evaluated in Gunrock, we focus in this paper on breadth-first
search (BFS), single-source shortest path (SSSP), betweenness
centrality (BC), PageRank, and connected components (CC). Though the
GPU's excellent peak throughput and energy
efficiency~\cite{Keckler:2011:GAT} have been demonstrated across many
application domains, these applications often exploit regular,
structured parallelism. The inherent irregularity of graph data
structures leads to irregularity in data access and control flow,
making an efficient implementation on GPUs a significant challenge.

Our goal with Gunrock is to deliver the performance of customized,
complex GPU hardwired graph primitives with a high-level programming
model that allows programmers to quickly develop new graph primitives.
To do so, we must address the chief challenge in a highly parallel
graph processing system: managing irregularity in work distribution.
Gunrock integrates sophisticated load-balancing and work-efficiency
strategies into its core. These strategies are hidden from the
programmer; the programmer instead expresses \emph{what} operations
should be performed on the frontier rather than \emph{how} those
operations should be performed. Programmers can assemble complex and
high-performance graph primitives from operations that manipulate the
frontier (the ``what'') without knowing the internals of the
operations (the ``how'').

Our contributions are as follows:

\begin{enumerate}
\item We present a novel data-centric abstraction for graph operations
  that allows programmers to develop graph primitives at a high level
  of abstraction while simultaneously delivering high performance.
  This abstraction, unlike the abstractions of previous GPU
  programmable frameworks, is able to elegantly incorporate profitable
  optimizations---kernel fusion, push-pull traversal, idempotent
  traversal, and priority queues---into the core of its
  implementation.
\item We design and implement a set of simple and flexible APIs that
  can express a wide range of graph processing primitives at a high
  level of abstraction (at least as simple, if not more so, than other
  programmable GPU frameworks).
\item We describe several GPU-specific optimization strategies for
  memory efficiency, load balancing, and workload management that
  together achieve high performance. All of our graph primitives
  achieve comparable performance to their hardwired counterparts and
  significantly outperform previous programmable GPU abstractions.
\item We provide a detailed experimental evaluation of our graph
  primitives with performance comparisons to several CPU and GPU
  implementations.
\end{enumerate}

Gunrock is currently available in an open-source repository at \emph{http://gunrock.github.io/} and is currently available for use by external developers.

\section{Related Work}
\label{sec:related}
This section discusses the research landscape of large-scale graph
analytics frameworks in four fields:

\begin{enumerate}
\item Single-node CPU-based systems, which are in common use for graph
    analytics today, but whose serial or coarse-grained-parallel programming
    models are poorly suited for a massively parallel processor like the GPU\@;
\item Distributed CPU-based systems, which offer scalability advantages over
    single-node systems but incur substantial communication cost, and whose
    programming models are also poorly suited to GPUs;
  \item GPU ``hardwired,'' low-level implementations of specific graph
    primitives, which provide a proof of concept that GPU-based graph
    analytics can deliver best-in-class performance. However,
    best-of-class hardwired primitives are challenging to even the
    most skilled programmers, and their implementations do not
    generalize well to a variety of graph primitives; and
  \item High-level GPU programming models for graph analytics, which
    often recapitulate CPU programming models (e.g., CuSha and
    MapGraph use PowerGraph's GAS programming model, Medusa uses
    Pregel's messaging model). The best of these systems incorporate
    generalized load balance strategies and optimized GPU primitives,
    but they generally do not compare favorably in performance with
    hardwired primitives due to the overheads inherent in a high-level
    framework and the lack of primitive-specific optimizations.
\end{enumerate}

\subsection{Single-node and Distributed CPU-based Systems} Parallel graph
analytics frameworks provide high-level, programmable, high-performance
abstractions.  The Boost Graph Library (BGL) is among the first efforts towards
this goal, though its serial formulation and C++ focus together make it poorly
suited for a massively parallel architecture like a GPU\@. Designed using the
generic programming paradigm, the parallel BGL~\cite{Gregor:2005:PBG} separates
the implementation of parallel algorithms from the underlying data structures
and communication mechanisms. While many BGL implementations are
specialized per algorithm, its breadth\_first\_visit pattern
(for instance) allows sharing common operators between different graph algorithms.
Pregel~\cite{Malewicz:2010:PSL} is Google's effort at large-scale graph
computing. It follows the Bulk Synchronous Parallel (BSP) model. A typical
application in Pregel is an iterative convergent process consisting of global
synchronization barriers called super-steps. The computation in Pregel is
vertex-centric and based on message passing. Its programming model is good for
scalability and fault tolerance. However, in standard graph algorithms in most
Pregel-like graph processing systems, slow convergence arises from graphs with
structure.  GraphLab~\cite{Low:2010:GAN} allows asynchronous computation and
dynamic asynchronous scheduling. By eliminating message-passing, its
programming model isolates the user-defined algorithm from the movement of
data, and therefore is more consistently expressive.
PowerGraph~\cite{Gonzalez:2012:PDG} uses the more flexible Gather-Apply-Scatter
(GAS) abstraction for power-law graphs. It supports both BSP and asynchronous
execution. For the load imbalance problem, it uses vertex-cut to split
high-degree vertices into equal degree-sized redundant vertices. This exposes
greater parallelism in natural graphs.  Ligra~\cite{Shun:2013:LAL} is
a CPU-based graph
processing framework for shared memory.  It uses a similar operator abstraction
for doing graph traversal. Its lightweight implementation is targeted at shared
memory architectures and uses CilkPlus for its multithreading implementation.
Galois~\cite{Pingali:2011:TTO, Nguyen:2013:ALI} is a graph system for shared
memory based on a different operator abstraction that supports priority
scheduling and dynamic graphs and processes on subsets of vertices called
active elements.  However, their model does not abstract the internal details
of the loop from the user. Users have to generate the active elements set
directly for different graph algorithms. Help is a library that
provides high-level primitives for large-scale graph
processing~\cite{Salihoglu:2014:HHP}.  Using the primitives in Help is more
intuitive and much faster than using the APIs of existing distributed systems.
Green-Marl~\cite{Hong:2012:GDE} is a domain-specific language for writing graph analysis algorithms on
shared memory with built-in breadth-first search (BFS) and depth-first search (DFS)
primitives in its compiler. Its language approach provides graph-specific optimizations
and hides complexity. However, the language does not support operations on
arbitrary sets of vertices for each iteration, which makes it difficult to use for
traversal algorithms that cannot be expressed using a BFS or DFS\@.

\subsection{Specialized Parallel Graph Algorithms} Recent work has developed
numerous best-of-breed, hardwired implementations of many graph primitives.
Merrill et al.~\cite{Merrill:2012:SGG}'s linear
parallelization of the BFS algorithm on the GPU had significant
influence in the field. They proposed an adaptive
strategy for load-balancing parallel work by expanding one node's neighbor list
to one thread, one warp, or a whole block of threads. With this strategy and
a memory-access efficient data representation, their implementation achieves
high throughput on large scale-free graphs. Beamer et al.'s recent work on
a very fast BFS for shared memory machines~\cite{Beamer:2012:DBS} uses a hybrid
BFS that switches between top-down and bottom-up neighbor-list-visiting
algorithms according to the size of the frontier to save redundant edge visits.
The current fastest connected-component algorithm on the GPU is Soman et al.'s
work~\cite{Soman:2010:AFG} based on two PRAM connected-component
algorithms~\cite{Greiner:1994:CPA}. There are several parallel Betweenness
Centrality implementations on the GPU~\cite{Geisberger:2008:BAO,
McLaughlin:2014:SAH, Pande:2011:CBC, Sariyuce:2013:BCO} based on the work from Brandes and
Ulrik~\cite{Brandes:2001:AFA}.  Davidson et al.~\cite{Davidson:2014:WPG:nourl}
proposed a work-efficient Single-Source Shortest Path algorithm on the GPU that
explores a variety of parallel load-balanced graph traversal and work
organization strategies to outperform other parallel methods. After we discuss
the Gunrock abstraction in Section~\ref{sec:frame:abstraction}, we will discuss
these existing hardwired GPU graph algorithm implementations using Gunrock
terminology.

\subsection{High-level GPU Programming Models} In
Medusa~\cite{Zhong:2014:MSG}, Zhong and He presented their pioneering
work on a high-level GPU-based system for parallel graph processing,
using a message-passing model. CuSha~\cite{Khorasani:2014:CVG},
targeting a GAS abstraction, implements the parallel-sliding-window
(PSW) graph representation on the GPU to avoid non-coalesced memory
access. CuSha additionally addresses irregular memory access by
preprocessing the graph data structure (``G-Shards'').
Both frameworks offer a small set of user-defined APIs but
are challenged by load imbalance and thus fail to achieve the same level of
performance as low-level GPU graph implementations.
MapGraph~\cite{Fu:2014:MAH} also adopts the GAS abstraction and
achieves some of the best performance results for programmable
single-node GPU graph computation.

\section{Background \& Preliminaries}
\label{sec:prelim}
A graph is an ordered pair $G=(V,E, w_e, w_v)$ comprised of a set of
vertices $V$ together with a set of edges $E$\@, where $E \subseteq V
\times V$. $w_e$ and $w_v$ are two weight functions that show the
weight values attached to edges and vertices in the graph. A graph is
undirected if for all $v,u \in V:$ $(v,u)\in E \Longleftrightarrow
(u,v)\in E$. Otherwise, it is directed. In graph processing, a vertex
frontier represents a subset of vertices $U\subseteq V$ and an edge frontier
represents a subset of edges $I\subseteq E$.

Modern NVIDIA GPUs are throughput-oriented manycore processors that
use massive parallelism to get very high peak computational throughput
and hide memory latency. Kepler-based GPUs can have up to 15 vector
processors, termed streaming multiprocessors (SMX), each containing 192
parallel processing cores, termed streaming processors (SP). NVIDIA
GPUs use the Single Instruction Multiple Thread (SIMT) programming
model to achieve data parallelism. GPU programs called \emph{kernels}
run on a large number of parallel threads. Each set of 32 threads forms a
divergent-free group called a \emph{warp} to execute in lockstep in a
Single Instruction Multiple Data (SIMD) fashion. These warps are then
grouped into cooperative thread arrays called \emph{blocks} whose
threads can communicate through a pool of on-chip shared memory. All
SMXs share an off-chip global DRAM\@.

For problems that require irregular data accesses such as graph
problems, in addition to exposing enough parallelism, a successful GPU
implementation benefits from the following application characteristics: 1) coalesced
memory access and effective use of the memory hierarchy, 2) minimizing
thread divergence within a warp, and 3) reducing scattered reads and
writes.

To achieve these goals, Gunrock represents all per-node and per-edge data
as structure-of-array (SOA) data structures that allow coalesced
memory accesses with minimal memory divergence. The data structure for
the graph itself is perhaps even more important. In Gunrock, we use a
compressed sparse row (CSR) sparse matrix for vertex-centric
operations by default and allow users to choose an edge-list-only representation
for edge-centric operations. CSR uses a column-indices array, $C$, to store
a list of neighbor vertices and a row-offsets array, $R$, to store the offset of the neighbor list for
each vertex. It provides compact and efficient memory access, and
allows us to use scan, a common and efficient parallel primitive, to
reorganize sparse and uneven workloads into dense and uniform ones in
all phases of graph processing~\cite{Merrill:2012:SGG}.

\section{The Gunrock Abstraction and Implementation}
\label{sec:frame}
\subsection{Gunrock's Abstraction}
\label{sec:frame:abstraction}

Gunrock targets graph operations that can be expressed as iterative
convergent processes. By ``iterative,'' we mean operations that may
require running a series of steps repeatedly; by ``convergent,'' we
mean that these iterations allow us to approach the correct answer and
terminate when that answer is reached. This target is similar to most
high-level graph frameworks.

Where Gunrock differs from other frameworks, particularly other
GPU-based frameworks, is in our abstraction. Rather than focusing on
sequencing steps of \emph{computation}, we instead focus on
manipulating a data structure, the \emph{frontier} of vertices or
edges that represents the subset of the graph that is actively
participating in the computation. It is accurate to say that for many
(but not all) computations, the sequence of operations that result from our
abstraction may be similar to what another abstraction may produce.
Nonetheless, we feel that thinking about graph processing in terms of
manipulations of frontier data structures is the right abstraction
for the GPU\@. We support this thesis qualitatively in this section and
quantitatively in Section~\ref{sec:performance-and-analysis}.

One important consequence of designing our abstraction with a
data-centered focus is that Gunrock, from its very beginning, has
supported both vertex and edge frontiers, and can easily switch
between them within the same graph primitive. We can, for instance,
generate a new frontier of neighboring edges from an existing frontier
of vertices. In contrast, gather-apply-scatter (PowerGraph/GAS) and
message-passing (Pregel) abstractions are focused on operations on
vertices and cannot easily support edge frontiers within their abstractions.

In our abstraction, we expose bulk-synchronous ``steps'' that
manipulate the frontier, and programmers build graph primitives from a
sequence of steps. Different steps may have dependencies between them,
but individual operations within a step can be processed in parallel.
For instance, a computation on each vertex within the frontier can be
parallelized across vertices, and updating the frontier by identifying
all the vertices neighboring the current frontier can also be
parallelized across vertices. BSP operations are well-suited for
efficient implementation on the GPU because they exhibit enough
parallelism to keep the GPU busy and do not require expensive
fine-grained synchronization or locking operations.

\begin{figure}[ht]
  \centering
  \includegraphics[width=0.45\textwidth]{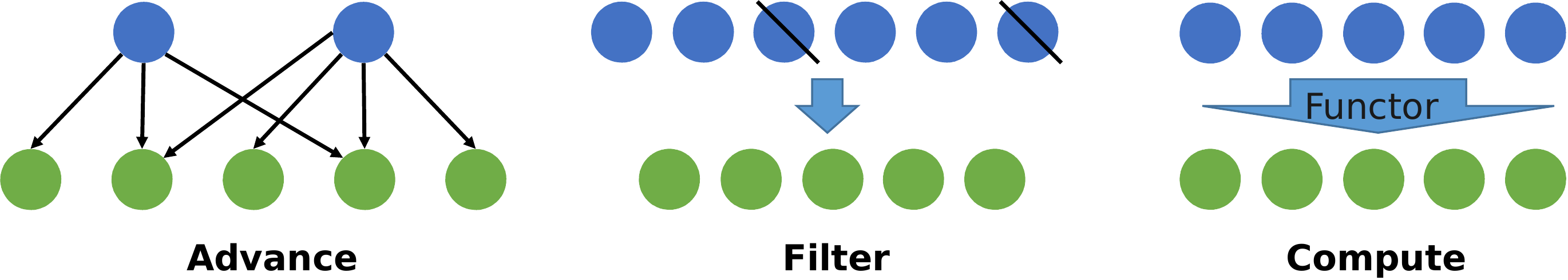}
  \centering
  \caption{Three operators in Gunrock's data-centric abstraction
    convert a current frontier (in blue) into a new frontier (in green).}
  \label{fig:data-centric}
\end{figure}

The graph primitives we describe in this paper use three Gunrock
steps---advance, filter, and compute---each of which manipulate
the frontier in a different way (Figure~\ref{fig:data-centric}).

\begin{description}[style=unboxed, leftmargin=0cm]

\item[Advance] An \emph{advance} step generates a new frontier from
  the current frontier by visiting the neighbors of the current
  frontier. A frontier can consist of either vertices or edges, and an
  advance step can input and output either
  kind of frontier. Advance is an irregularly-parallel operation for
  two reasons: (1) different vertices in a graph have different
  numbers of neighbors and (2) vertices share neighbors, so an
  efficient advance is the most significant challenge of a GPU
  implementation.

  The generality of Gunrock's advance allows us to use the same
  advance implementation across a wide variety of interesting graph
  operations. For instance, we can utilize Gunrock advance operators to: 1)~visit
  each element in the current frontier while updating local values
  and/or accumulating global values (e.g., BFS distance updates); 2)~visit the vertex or edge
  neighbors of all the elements in the current frontier while updating
  source vertex, destination vertex, and/or edge values (e.g., distance updates in SSSP); 3)~generate
  edge frontiers from vertex frontiers or vice versa (e.g., BFS, SSSP, depth-first search, etc.); or 4)~pull
  values from all vertices 2 hops away by starting from an edge
  frontier, visiting all the neighbor edges, and returning the far-end
  vertices of these neighbor edges. As a result, we can concentrate
  our effort on solving one problem (implementing an efficient
  advance) and see that effort reflected in better performance on
  other traversal-based graph operations.

\item[Filter] A \emph{filter} step generates a new frontier from the
  current frontier by choosing a subset of the current frontier based
  on programmer-specified criteria. Though filtering is an irregular
  operation, using parallel scan for efficient filtering is
  well-understood on GPUs. Gunrock's filters can either 1)~split
  vertices or edges based on a filter (e.g., SSSP's delta-stepping),
  or 2)~compact out filtered items to throw them away (e.g., duplicate
  vertices in BFS, SSSP, and BC).

\item[Compute] A programmer-specified \emph{compute} step
  defines an operation on all elements (vertices or edges) in the
  current frontier; Gunrock then performs that operation in parallel
  across all elements. Because this parallelism is regular,
  computation is straightforward to parallelize in a GPU
  implementation. Many simple graph primitives (e.g., computing the
  degree distribution of a graph) can be expressed as a single
  computation step.
\end{description}

\noindent
Gunrock primitives are assembled from a sequence of these steps, which
are executed sequentially: one step completes all of its operations
before the next step begins. Typically, Gunrock graph primitives run
to convergence, which on Gunrock usually equates to an empty frontier;
as individual elements in the current frontier reach convergence, they
can be filtered out of the frontier. Programmers can also use other
convergence criteria such as a maximum number of iterations or
volatile flag values that can be set in a computation step.

\paragraph{Expressing SSSP in programmable GPU frameworks}
SSSP is a reasonably complex graph primitive that computes the shortest path
from a single node in a graph to every other node in the graph. We assume
weights between nodes are all non-negative, which permits the use of Dijkstra's
algorithm and its parallel variants. Efficiently implementing SSSP continues to
be an interesting problem in the GPU
world~\cite{Burtscher:2012:AQS,Davidson:2014:WPG:nourl,Delling:2010:PHS}.

The iteration starts with an input frontier of
active vertices (or a single vertex) initialized to a distance of zero.
First, SSSP enumerates the sizes of the frontier's neighbor
list of edges and computes the length of the output frontier. Because the
neighbor edges are unequally distributed among the frontier's vertices, SSSP
next redistributes the workload across parallel threads. This can be expressed
within an advance frontier. In the final step of the advance frontier, each edge adds
its weight to the distance value at its source value and, if appropriate,
updates the distance value of its destination vertex.  Finally, SSSP removes
redundant vertex IDs (specific filter), decides which updated vertices are valid in the new
frontier, and computes the new frontier for the next iteration.

  Algorithm~\ref{ssspcode} provides more detail of how this algorithm
  maps to Gunrock's abstraction.

  \begin{algorithm}[!ht] \caption{Single-Source Shortest Path, expressed
      in Gunrock's abstraction} \label{ssspcode}
    \begin{small}
      \begin{algorithmic}[1]
        \Procedure{Set\_Problem\_Data}{$G, P, root$}
        \State $P.\textit{labels}[1..G.\textit{verts}] \gets \infty$
        \State $P.\textit{preds}[1..G.\textit{verts}] \gets -1$
        \State $P.\textit{labels}[\textit{root}] \gets 0$
        \State $P.\textit{preds}[\textit{root}] \gets \textit{src}$
        \State $P.\textit{frontier}.\textit{Insert}(\textit{root})$
        \EndProcedure
        \State
        \Procedure{UpdateLabel}{$s\_id, d\_id, e\_id, P$}
        \State $\textit{new}\_\textit{label} \gets P.\textit{labels}[s\_id]+P.\textit{weights}[e\_id]$
        \State \textbf{return} $\textit{new}\_\textit{label} < \textit{atomicMin}(P.\textit{labels}[d\_id], \textit{new}\_\textit{label})$
        \EndProcedure
        \State
        \Procedure{SetPred}{$s\_id, d\_id, P$}
        \State $P.\textit{preds}[d\_id] \gets s\_id$
        \State $P.\textit{output}\_\textit{queue}\_\textit{ids}[d\_id] \gets \textit{output}\_\textit{queue}\_\textit{id}$
        \EndProcedure
        \State
        \Procedure{RemoveRedundant}{$\textit{node}\_\textit{id},P$}
        \State \textbf{return} $P.\textit{output}\_\textit{queue}\_\textit{id}[\textit{node}\_\textit{id}] == \textit{output}\_\textit{queue}\_\textit{id}$
        \EndProcedure
        \State
        \Procedure{SSSP\_Enactor}{$G,P,\textit{root}$}
        \State
        \Call {Set\_Problem\_Data}{$G,P,\textit{root}$}
        \While{$P.\textit{frontier}.\textit{Size}() > 0$}
        \State
        \Call {Advance}{$G,P,\textit{UpdateLabel},\textit{SetPred}$}
        \State
        \Call {Filter}{$G,P,\textit{RemoveRedundant}$}
        \State
        \Call {PriorityQueue}{$G,P$}
        \EndWhile
        \EndProcedure
      \end{algorithmic}
    \end{small}
  \end{algorithm}

Gunrock maps one SSSP iteration onto three Gunrock steps: (1)
\emph{advance}, which computes the list of edges connected to the
current vertex frontier and (transparently) load-balances their
execution; (2) \emph{compute}, to update neighboring vertices with new
distances; and (3) \emph{filter}, to generate the final output
frontier by removing redundant nodes, optionally using a 2-level
priority queue, whose use enables delta-stepping (a binning strategy
to reduce overall workload~\cite{Davidson:2014:WPG:nourl,Meyer:2003:DAP}).
With this mapping in place, the traversal and computation of path distances
is simple and intuitively described, and Gunrock is able to create an efficient
implementation that fully utilizes the GPU's computing resources in a load-balanced way.

\begin{figure*}[ht]
    \centering
    \includegraphics[width=0.9\textwidth]{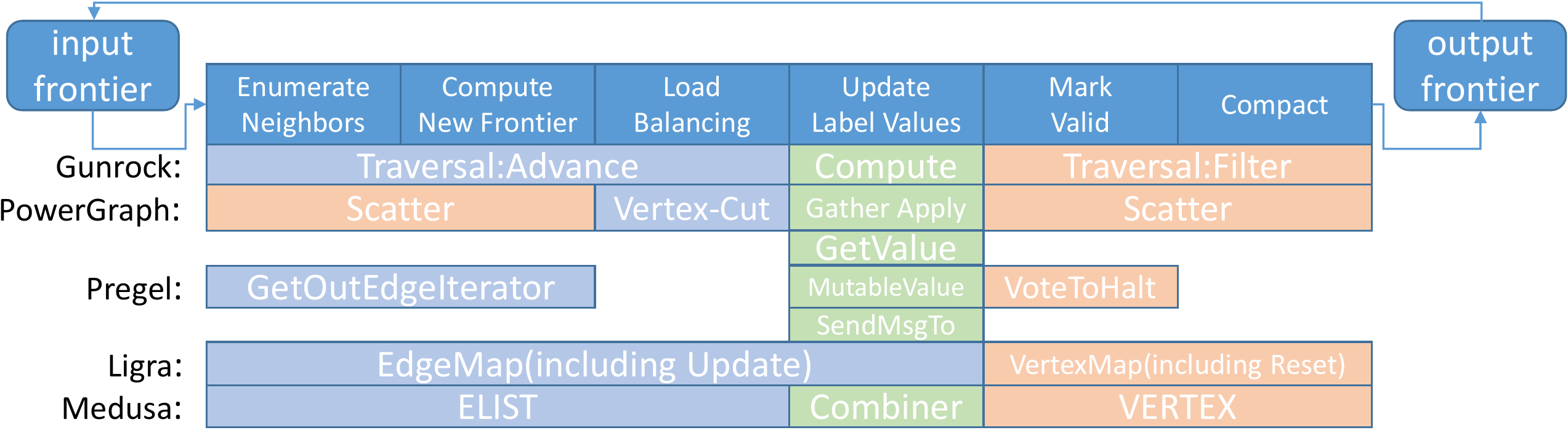}
    \centering
    \caption{Operations that make up one iteration of SSSP and their mapping to the Gunrock, PowerGraph (GAS)~\protect\cite{Gonzalez:2012:PDG}, Pregel~\protect\cite{Malewicz:2010:PSL}, Ligra~\protect\cite{Shun:2013:LAL}, and Medusa~\protect\cite{Zhong:2014:MSG} abstractions.}
    \label{fig:abstraction}
\end{figure*}

\subsection{Alternative Abstractions}
In this section we discuss several alternative abstractions designed for graph
processing on various architectures.

\begin{description}[style=unboxed, leftmargin=0cm]
\item[Gather-apply-scatter (GAS) abstraction]
  The GAS abstraction was first applied on distributed
  systems~\cite{Gonzalez:2012:PDG}. PowerGraph's vertex-cut splits large
  neighbor lists, duplicates node information, and deploys each partial
  neighbor list to different machines. Working as a load balancing strategy,
  it replaces the large synchronization cost in edge-cut into a single-node
  synchronization cost. This is a productive strategy for multi-node
  implementations. GAS abstractions have successfully been mapped to the GPU,
  first with VertexAPI2~\cite{Elsen:2013:AVC:url} and later with
  MapGraph~\cite{Fu:2014:MAH} and CuSha~\cite{Khorasani:2014:CVG}. GAS offers
  the twin benefits of simplicity and familiarity, given its popularity in
  the CPU world.

  Recently, Wu et al. compared Gunrock vs.\ two GPU GAS
  frameworks, VertexAPI2 and MapGraph~\cite{Wu:2015:PCF},
  demonstrating that Gunrock had appreciable performance advantages
  over the other two frameworks. One of the principal performance
  differences they identified comes from the significant fragmentation
  of GAS programs across many kernels that we discuss in more detail in
  Section~\ref{sec:gunrock-kernel-fusion}. Applying automatic kernel
  fusion to GAS+GPU implementations could potentially help close their
  performance gap, but such an optimization is highly complex and has
  not yet appeared in any published work.

  At a more fundamental level, we found that a compute-focused
  programming model like GAS was not flexible enough to manipulate the
  core frontier data structures in a way that enabled powerful
  features and optimizations such as push-pull and two-level priority queues;
  both fit naturally into Gunrock's abstraction. We believe bulk-synchronous
  operations on frontiers are a better fit than GAS for forward-looking GPU
  graph programming frameworks.

\item[Message-passing] Pregel~\cite{Malewicz:2010:PSL} is a
  vertex-centric programming model that only provides data parallelism
  on vertices. For graphs with significant variance in vertex degree
  (e.g., power-law graphs), this would cause severe load imbalance on
  GPUs. The traversal operator in Pregel is general enough to apply to
  a wide range of graph primitives, but its vertex-centric design only
  achieves good parallelism when nodes in the graph have small and
  evenly-distributed neighborhoods. For real-world graphs that often
  have uneven distribution of node degrees, Pregel suffers from severe
  load imbalance. The Medusa GPU graph-processing
  framework~\cite{Zhong:2014:MSG} also implements a BSP model and
  allows computation on both edges and vertices. Medusa, unlike
  Gunrock, also allows edges and vertices to send \emph{messages} to
  neighboring vertices. The Medusa authors note the complexity of
  managing the storage and buffering of these messages, and the
  difficulty of load-balancing when using segmented reduction for
  per-edge computation. Though they address both of these challenges
  in their work, the overhead of \emph{any} management of messages is
  a significant contributor to runtime. Gunrock prefers the less
  costly direct communication between primitives and supports both
  push-based (scatter) communication and pull-based (gather)
  communication during traversal steps.

\item[CPU strategies] Ligra's powerful load-balancing strategy is
  based on CilkPlus, a fine-grained task-parallel library for CPUs.
  Despite promising GPU research efforts on task
  parallelism~\cite{Cederman:2008:ODL,Tzeng:2012:AGT:nourl}, no such
  equivalent is available on GPUs, thus we implement our own
  load-balancing strategies within Gunrock. Galois, like Gunrock,
  cleanly separates data structures from computation; their key
  abstractions are ordered and unordered set iterators that can add
  elements to sets during execution (such a dynamic data structure is
  a significant research challenge on GPUs). Galois also benefits from
  speculative parallel execution whose GPU implementation would also
  present a significant challenge. Both Ligra and Galois scale well
  within a node through inter-CPU shared memory; inter-GPU
  scalability, both due to higher latency and a lack of hardware
  support, is a much more manual, complex process.

\item[Help's Primitives] Help~\cite{Salihoglu:2014:HHP}
  characterizes graph primitives as a set of functions that enable
  special optimizations for different primitives at the cost of
  losing generality. Its Filter, Local Update of Vertices (LUV),
  Update Vertices Using One Other Vertex (UVUOV), and Aggregate
  Global Value (AGV) are all Gunrock filter operations with
  different computations. Aggregating Neighbor Values (ANV) maps to
  the advance operator in Gunrock. We also successfully implemented
  FS in Gunrock using two filter passes, one advance pass, and
  several other GPU computing primitives (sort, reduce, and scan).

\item[Asynchronous execution] Many CPU frameworks (e.g., Galois and
  GraphLab) efficiently incorporate asynchronous execution, but the
  GPU's expensive synchronization or locking operations would make
  this a poor choice for Gunrock. We do recover some of the benefits
  of prioritizing execution through our two-level priority queue.

\end{description}

\subsection{Gunrock's API and its Kernel-Fusion Optimization}
\label{sec:gunrock-kernel-fusion}
\begin{figure}[ht]
    \centering
    \includegraphics[width=0.45\textwidth]{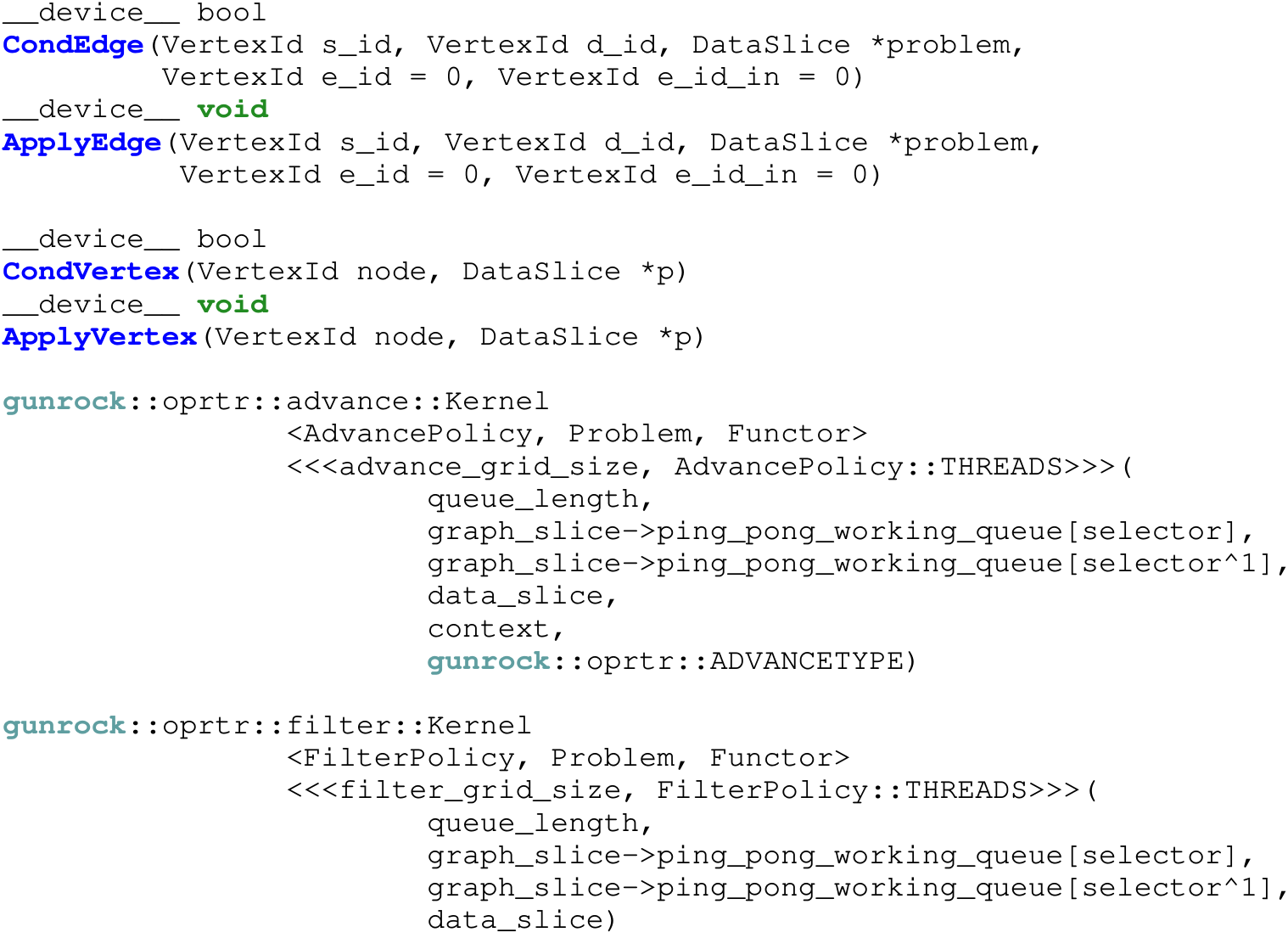}
  \caption{Gunrock's API set. \texttt{Cond} functors compute a boolean
    value per element, useful for filtering. \texttt{Apply} functors
    implement a compute operation on each element. User specific functor
    struct that contains its own implementation of these four functors
    is integrated at compile time into \texttt{Advance} or
    \texttt{Filter} kernels, providing automatic kernel
    fusion.}\label{fig:apiset}
\end{figure}

Gunrock programs specify three components: the \emph{problem}, which
provides graph topology data and an algorithm-specific data management
interface; the \emph{functors}, which contain user-defined computation
code and expose kernel fusion opportunities that we discuss below; and
an \emph{enactor}, which serves as the entry point of the graph
algorithm and specifies the computation as a series of advance and/or
filter kernel calls with user-defined kernel launching settings.

Given Gunrock's abstraction, the most natural way to specify Gunrock
programs would be as a sequence of bulk-synchronous steps, specified
within the enactor and implemented as kernels, that operate on
frontiers. Such an enactor is in fact the core of a Gunrock program,
but an enactor-only program would sacrifice a significant performance
opportunity. We analyzed the techniques that hardwired
(primitive-specific) GPU graph primitives used to achieve high
performance. One of their principal advantages is leveraging
producer-consumer locality between operations by integrating multiple
operations into single GPU kernels. Because adjacent kernels in CUDA
or OpenCL share no state, combining multiple logical operations into a
single kernel saves significant memory bandwidth that would otherwise
be required to write and then read intermediate values to and from
memory. The CUDA C++ programming environment we use has no ability to
automatically fuse neighboring kernels together to achieve this
efficiency (and automating this ``kernel fusion'' problem is a
significant research challenge).

In particular, we noted that hardwired GPU implementations fuse
regular computation steps together with more irregular steps like
advance and filter by running a computation step (with regular
parallelism) on the input or output of the irregularly-parallel step,
all within the same kernel. To enable similar behavior in a
programmable way, Gunrock exposes its computation steps as
\emph{functors} that are integrated into advance and filter kernels at
compile time to achieve similar efficiency. We support functors that
apply to \{edges, vertices\} and either return a boolean value (the
``cond'' functor), useful for filtering, or perform a computation (the
``apply'' functor). These functors will then be integrated into
``advance'' and ``filter'' kernel calls, which hide any complexities
of how those steps are internally implemented.
We summarize the API for these operations in Figure~\ref{fig:apiset}.
Our focus on kernel fusion
enabled by our API design is absent from other programmable GPU graph
libraries, but it is crucial for performance.

In terms of data structures, Gunrock represents all per-node and per-edge data
as structure-of-array (SOA) data structures that allow coalesced
memory accesses with minimal memory divergence. The data structure for
the graph itself is perhaps even more important. In Gunrock, we use a
compressed sparse row (CSR) sparse matrix for vertex-centric
operations by default and allow users to choose an edge-list-only representation
for edge-centric operations. CSR uses a column-indices array, $C$, to store
a list of neighbor vertices and a row-offsets array, $R$, to store the offset of the neighbor list for
each vertex. It provides compact and efficient memory access, and
allows us to use scan, a common and efficient parallel primitive, to
reorganize sparse and uneven workloads into dense and uniform ones in
all phases of graph processing~\cite{Merrill:2012:SGG}.

We next provide detail on Gunrock's implementations of
workload-mapping/load-balancing (Section~\ref{sec:workload-mapping})
and optimizations (Section~\ref{sec:optimizations})

\subsection{Workload Mapping and Load Balancing Details}
\label{sec:frame:optimize}
\label{sec:workload-mapping}
\label{sec:work-efficiency}

Choosing the right abstraction is one key component in achieving high
performance within a graph framework. The second component is
optimized implementations of the primitives within the framework. One
of Gunrock's major contributions is generalizing two workload-distribution
and load-balance strategies that each previously applied to a single
hardwired GPU graph primitive into Gunrock's general-purpose advance
operator.

Gunrock's advance step generates an irregular workload. Consider an
advance that generates a new vertex frontier from the neighbors of all
vertices in the current frontier. If we parallelize over input
vertices, graphs with a variation in vertex degree (with
different-sized neighbor lists) will generate a corresponding
imbalance in per-vertex work. Thus, mapping the workload of each
vertex onto the GPU so that they can be processed in a load-balanced
way is essential for efficiency.

The most significant previous work in this area balances load by
cooperating between threads. Targeting BFS, Merrill et
al.~\cite{Merrill:2012:SGG} map the workload of a single vertex to a
thread, a warp, or a cooperative thread array (CTA), according to the
size of its neighbor list. Targeting SSSP, Davidson et
al.~\cite{Davidson:2014:WPG:nourl} use two load-balanced workload
mapping strategies, one that groups input work and the other that
groups output work. The first partitions the frontier into equally
sized chunks and assigns all neighbor lists of one chunk to one block;
the second partitions the neighbor list set into equally sized chunks
(possibly splitting the neighbor list of one node into multiple
chunks) and assigns each chunk of edge lists to one block of threads.
Merrill et al.\ (unlike Davidson et al.)
also supports the (BFS-specific) ability to process frontiers of edges
rather than just frontiers of vertices. We integrate both techniques
together, generalize them into a generic advance operator, and extend
them by supporting an effective pull-based optimization strategy
(Section~\ref{sec:optimizations}). The result is the following two
load-balancing strategies within Gunrock.

\begin{description}[style=unboxed, leftmargin=0cm]
\item[Per-thread fine-grained] One straightforward approach to load
  balancing is to map one frontier vertex's neighbor list to one
  thread. Each thread loads the neighbor list offset for its assigned
  node, then serially processes edges in its neighbor list. We have
  improved this method in several ways. First, we load all the
  neighbor list offsets into shared memory, then use a CTA of threads
  to cooperatively process per-edge operations on the neighbor list.
  Simultaneously, we use vertex-cut to split the neighbor list of a
  node so that it can be processed by multiple threads. We found out
  that this method performs better when used for large-diameter graphs
  with a relatively even degree distribution since it balances thread
  work within a CTA, but not across CTAs. For graphs with a more
  uneven degree distribution (e.g., scale-free social graphs), we turn
  to a second strategy.

\item[Per-warp and per-CTA coarse-grained] Significant differences in
  neighbor list size cause the worst performance with our per-thread
  fine-grained strategy. We directly address the variation in size by
  grouping neighbor lists into three categories based on their size,
  then individually processing each category with a strategy targeted
  directly at that size. Our three sizes are (1) lists larger than a
  CTA; (2) lists larger than a warp (32 threads) but smaller than a
  CTA; and (3) lists smaller than a warp. We begin by assigning a
  subset of the frontier to a block. Within that block, each thread
  owns one node. The threads that own nodes with large lists arbitrate
  for control of the entire block. All the threads in the block then
  cooperatively process the neighbor list of the winner's node. This
  procedure continues until all nodes with large lists have been
  processed. Next, all threads in each warp begin a similar procedure
  to process all the nodes whose neighbor lists are medium-sized
  lists. Finally, the remaining nodes are processed using our
  per-thread fine-grained workload-mapping strategy
  (Figure~\ref{fig:workload1}).

  \begin{figure*}[ht]
    \begin{minipage}[b]{0.5\linewidth}
      \centering
      \includegraphics[width=0.7\textwidth]{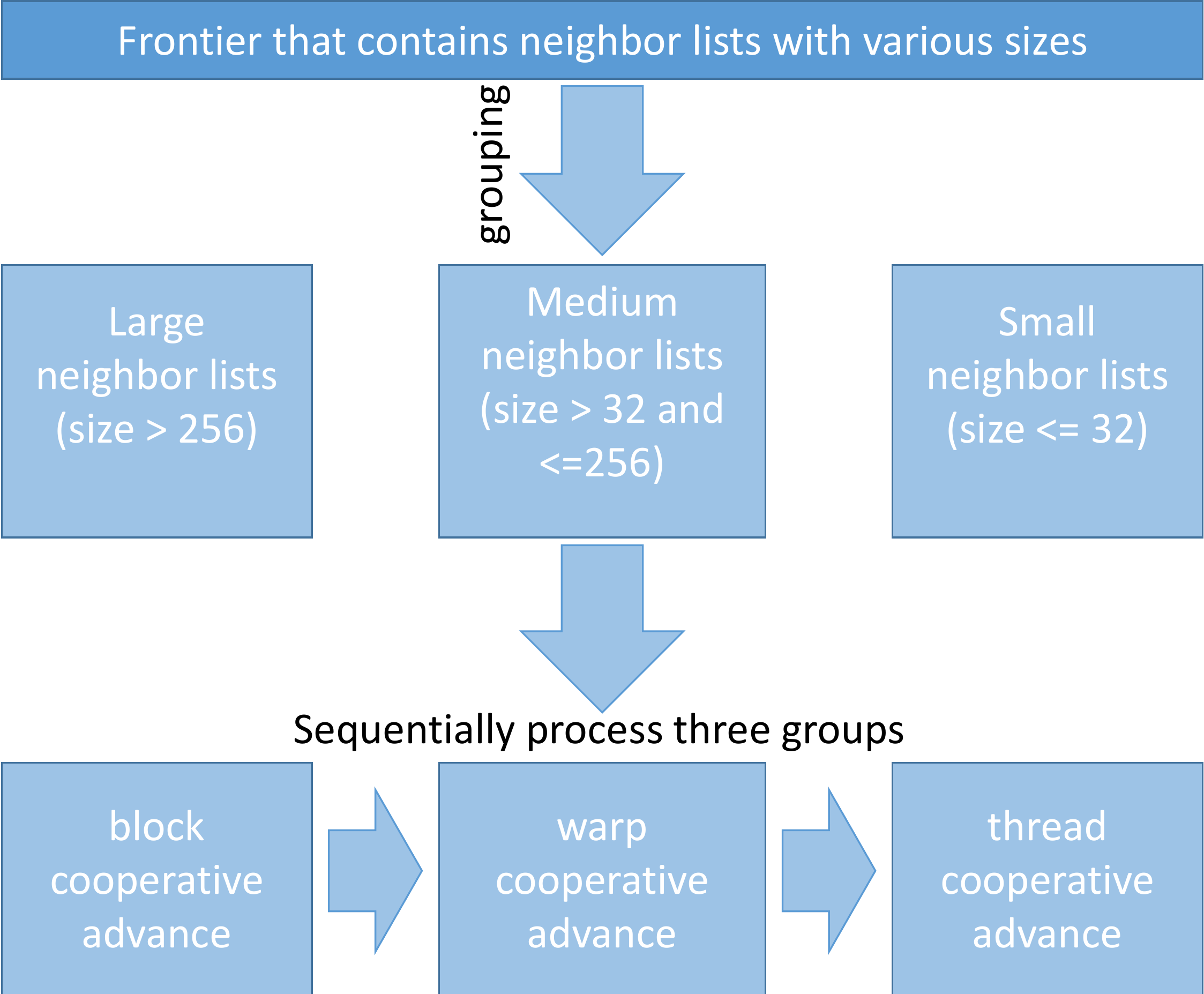}
      \caption{Load balancing strategy of Merrill et  al.~\cite{Merrill:2012:SGG}}
      \label{fig:workload1}
    \end{minipage}
    \hspace{0.1cm}
    \begin{minipage}[b]{0.5\linewidth}
      \centering
      \includegraphics[width=0.7\textwidth]{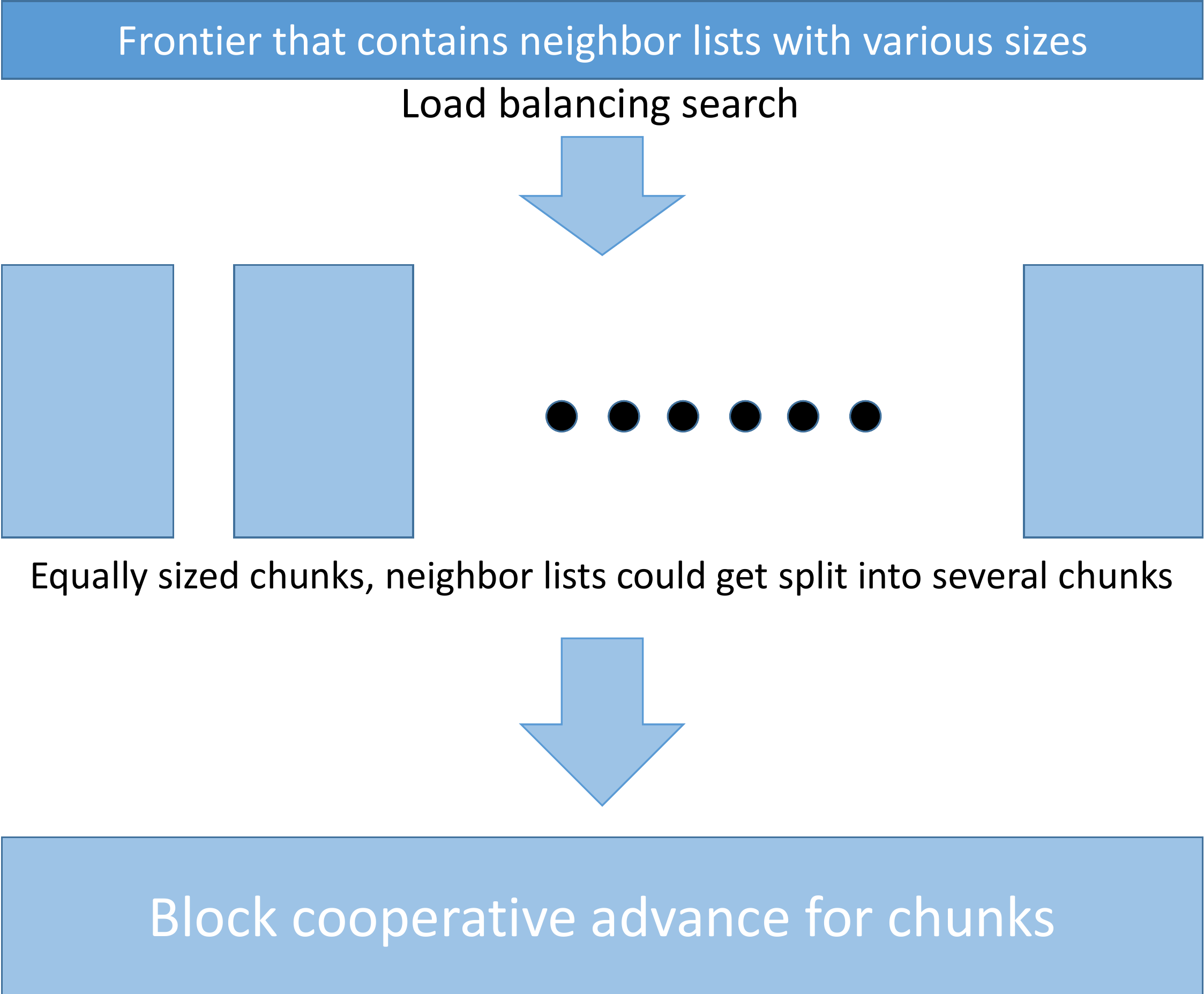}
      \caption{Load balancing strategy of Davidson et al.~\cite{Davidson:2014:WPG:nourl}}
      \label{fig:workload2}
    \end{minipage}
  \end{figure*}

  The specialization of this method allows higher throughput on
  frontiers with a high variance in degree distribution, but at the
  cost of higher overhead due to the sequential processing of the
  three different sizes.

\item[Load-Balanced Partitioning]
  Davidson et al.\ and Gunrock improve on this
  method by first organizing groups of edges into equal-length chunks
  and assigning each chunk to a block. This division requires us to
  find the starting and ending indices for all the blocks within the
  frontier. We use an efficient sorted search to map such indices with
  the scanned edge offset queue. When we start to process a neighbor
  list of a new node, we use binary search to find the node ID for the
  edges that are going to be processed. Using this method, we ensure
  load-balance both within a block and between blocks
  (Figure~\ref{fig:workload2}).

  At the high level, Gunrock makes a load-balancing strategy decision depending
  on topology. We note that our coarse-grained (load-balancing) traversal
  method performs better on social graphs with irregular distributed degrees,
  while the fine-grained method is superior on graphs where most nodes have
  small degrees. For this reason, in Gunrock we implement a hybrid of both
  methods on both vertex and edge frontiers, using the fine-grained dynamic
  grouping strategy for nodes with relatively smaller neighbor lists and the
  coarse-grained load-balancing strategy for nodes with relatively larger
  neighbor lists. Within the latter, we set a static threshold. When the frontier
  size is smaller than the threshold, we use coarse-grained load-balance over
  nodes, otherwise coarse-grained load-balance over edges. We have found that
  setting this threshold to 4096 yields consistent high performance for tests
  across all Gunrock-provided graph primitives. Users can also change this
  value easily in the Enactor module for their own datasets or graph
  primitives. Superior load balancing is one of the most significant reasons
  why Gunrock outperforms other GPU frameworks~\cite{Wu:2015:PCF}.
\end{description}

\subsection{Gunrock's Optimizations}
\label{sec:optimizations}

One of our main goals in designing the Gunrock abstraction was to
easily allow integrating existing and new alternatives and
optimizations into our primitives to give more options to programmers.
In general, we have found that our data-centric abstraction, and our
focus on manipulating the frontier, has been an excellent fit for
these alternatives and optimizations, compared to a more difficult
implementation path for other GPU computation-focused abstractions. We
offer three examples.

\begin{description}[style=unboxed, leftmargin=0cm]

\item[Idempotent vs.\ non-idempotent operations] Because multiple
  elements in the frontier may share a common neighbor, an advance
  step may generate an output frontier that has duplicated elements.
  For some graph primitives (e.g., BFS) with ``idempotent''
  operations, repeating a computation causes no harm, and Gunrock's
  filter step can perform a series of inexpensive heuristics to
  reduce, but not eliminate, redundant entries in the output frontier.
  Gunrock also supports a non-idempotent advance, which internally
  uses atomic operations to guarantee each element appears only once
  in the output frontier.

\item[Push vs.\ pull traversal] Other GPU programmable graph
  frameworks also support an advance step, of course, but because they
  are centered on vertex operations on an implicit frontier, they
  generally support only ``push''-style advance: the current frontier
  of active vertices ``pushes'' active status to its neighbors to
  create the new frontier. Beamer et al.~\cite{Beamer:2012:DBS}
  described a ``pull''-style advance on CPUs: instead of starting with
  a frontier of active vertices, pull starts with a frontier of
  \emph{unvisited} vertices, generating the new frontier by filtering
  the unvisited frontier for vertices that have neighbors in the
  current frontier.

  Beamer et al.\ showed this approach is beneficial
  when the number of unvisited vertices drops below the size of the
  current frontier. While vertex-centered GPU frameworks have found it
  challenging to integrate this optimization into their abstraction,
  our data-centric abstraction is a much more natural fit because we
  can easily perform more flexible operations on frontiers. Gunrock
  internally converts the current frontier into a bitmap of vertices,
  generates a new frontier of all unvisited nodes, then uses an
  advance step to ``pull'' the computation from these nodes'
  predecessors if they are valid in the bitmap.

  With this optimization, we see a speedup on BFS of 1.52x for
  scale-free graphs and 1.28x for small-degree-large-diameter graphs.
  In an abstraction like Medusa, with its fixed method (segmented
  reduction) to construct frontiers, it would be a significant
  challenge to integrate a pull-based advance. Currently in Gunrock,
  this optimization is applied to BFS only, but in the future, more
  sophisticated BC and SSSP implementations could benefit from it as
  well.

\item[Priority Queue] A straightforward BSP implementation of an
  operation on a frontier treats each element in the frontier equally,
  i.e., with the same priority. Many graph primitives benefit from
  prioritizing certain elements for computation with the expectation
  that computing those elements first will save work overall (e.g.,
  delta-stepping for SSSP~\cite{Meyer:2003:DAP}). Gunrock generalizes
  the approach of Davidson et al.~\cite{Davidson:2014:WPG:nourl} by
  allowing user-defined priority functions to organize an output
  frontier into ``near'' and ``far'' slices. This allows the GPU to
  use a simple and high-performance split operation to create and
  maintain the two slices. Gunrock then considers only the near slice
  in the next processing steps, adding any new elements that do not
  pass the near criterion into the far slice, until the near slice is
  exhausted. We then update the priority function and operate on the
  far slice.

  Like other Gunrock steps, constructing a priority queue directly
  manipulates the frontier data structure. It is difficult to
  implement such an operation in a GAS-based programming model since
  that programming model has no explicit way to reorganize a frontier.

  Currently Gunrock uses this specific optimization only in SSSP, but
  we believe a workload reorganization strategy based on a more
  general priority queue implementation will enable a
  semi-asynchronous execution model in Gunrock since different parts of
  frontier can process an arbitrary number of BSP steps. This will potentially
  increase the performance of various types of community detection and
  label propagation algorithms as well as algorithms on graphs with
  small ``long tail'' frontiers.

\end{description}

\section{Applications}
\label{sec:app}
One of the principal advantages of Gunrock's abstraction is that our
advance, filter, and compute steps can be composed to build new graph
primitives with minimal extra work. For each primitive below, we
describe the hardwired GPU implementation to which we compare,
followed by how we express this primitive in Gunrock.
Section~\ref{sec:performance-and-analysis} compares the performance
between hardwired and Gunrock implementations.

\begin{figure}[htb]
  \centering
  \includegraphics[width=\columnwidth]{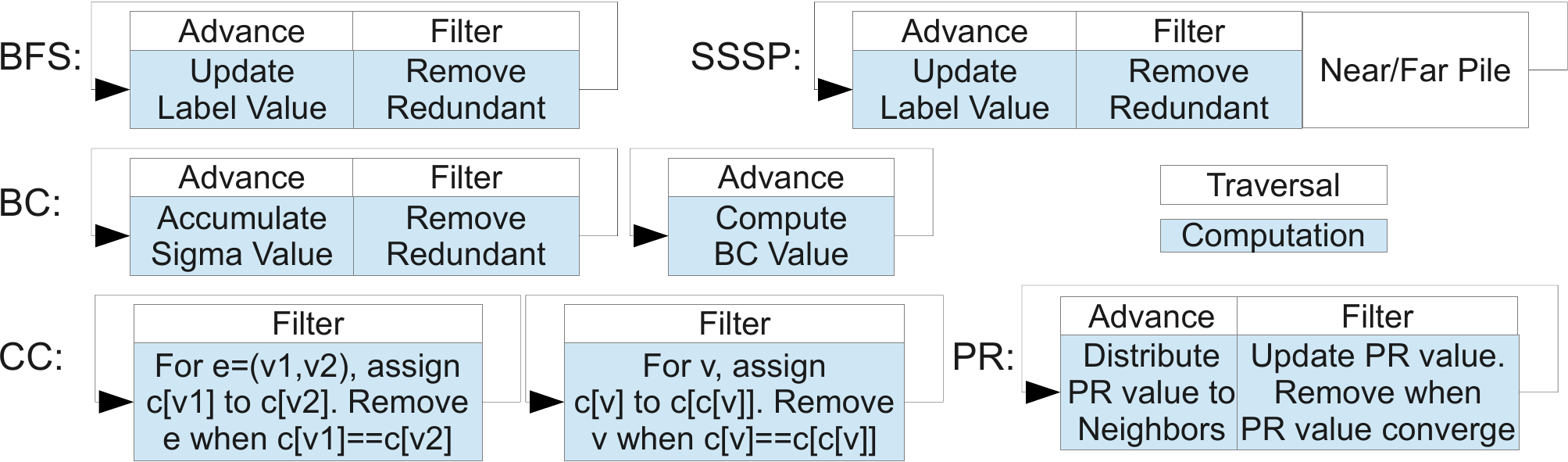}
  \centering
  \caption{Operation flow chart for selected primitives in Gunrock
    (a black line with an arrow at one end indicates a while loop
    that runs until the frontier is empty).\label{fig:flow}}
\end{figure}

\subsection{Breadth-First Search (BFS)}

BFS initializes its vertex frontier with a single source vertex. On
each iteration, it generates a new frontier of vertices with all
unvisited neighbor vertices in the current frontier, setting their
depths and repeating until all vertices have been visited. BFS is one
of the most fundamental graph primitives and serves as the basis of
several other graph primitives.

\begin{description}[style=unboxed, leftmargin=0cm]

\item[Hardwired GPU Implementation] The well-known BFS implementation
  of Merrill et al.~\cite{Merrill:2012:SGG} achieves its high
  performance through careful load-balancing, avoidance of atomics,
  and heuristics for avoiding redundant vertex discovery. Its chief
  operations are expand (to generate a new frontier) and contract (to
  remove redundant vertices) phases.

\item[Gunrock Implementation] Merrill et al.'s expand maps nicely to
  Gunrock's advance operator, and contract to Gunrock's filter
  operator. During advance, we set a label value for each vertex to
  show the distance from the source, and/or set a predecessor value
  for each vertex that shows the predecessor vertex's ID\@. We
  implement efficient load-balancing
  (Section~\ref{sec:workload-mapping}) and both push- and pull-based
  advance (Section~\ref{sec:optimizations}) for more efficient
  traversal. Our base implementation uses atomics during advance to prevent concurrent
  vertex discovery. When a vertex is uniquely discovered, we set its
  label (depth) and/or predecessor ID\@. Gunrock's fastest BFS uses the
  idempotent advance operator (thus avoiding the cost of atomics) and
  uses heuristics within its filter that reduce the concurrent
  discovery of child nodes (Section~\ref{sec:optimizations}).

\end{description}

\subsection{Single-Source Shortest Path}
Single-source shortest path finds paths between a given source vertex
and all other vertices in the graph such that the weights on the path
between source and destination vertices are minimized. While the
advance mode of SSSP is identical to BFS, the computation mode
differs.

\begin{description}[style=unboxed, leftmargin=0cm]
\item[Hardwired GPU Implementation] Currently the highest performing
  SSSP algorithm implementation on the GPU is the work from Davidson
  et al.~\cite{Davidson:2014:WPG:nourl}. They provide two key
  optimizations in their SSSP implementation: 1)~a load balanced graph
  traversal method and 2)~a priority queue implementation that
  reorganizes the workload. Gunrock generalizes both optimization
  strategies into its implementation, allowing them to apply to other
  graph primitives as well as SSSP\@. We implement Gunrock's priority
  queue as an additional filter pass between two iterations.

\item[Gunrock Implementation] We start from a single source vertex in
  the frontier. To compute a distance value from the source vertex, we
  need one advance and one filter operator. On each iteration, we
  visit all associated edges in parallel for each vertex in the
  frontier and relax the distance's value (if necessary) of the
  vertices attached to those edges. We use an AtomicMin to atomically
  find the minimal distance value we want to keep and a bitmap flag
  array associated with the frontier to remove redundant vertices.
  After each iteration, we use a priority queue to reorganize the
  vertices in the frontier.
\end{description}

\subsection{Betweenness Centrality}
\label{sec:app:bc}

The BC index can be used in social network analysis as an indicator of
the relative importance of vertices in a graph. At a high level, the
BC for a vertex in a graph is the fraction of shortest paths in a
graph that pass through that vertex. Brandes's BC
formulation~\cite{Brandes:2001:AFA} is most commonly used for GPU
implementations.

\begin{description}[style=unboxed, leftmargin=0cm]
\item[Hardwired GPU Implementation] Brandes's formulation has two
  passes: a forward BFS pass to accumulate sigma values for each node,
  and a backward BFS pass to compute centrality values. Jia et
  al.~\cite{Jia:2011:ENP} and Sariy\"{u}ce et
  al.~\cite{Sariyuce:2013:BCO} both use an edge-parallel method to
  implement the above two passes. We achieve this in Gunrock using two
  advance operators on an edge frontier with different computations.
  The recent (hardwired) multi-GPU BC algorithm by McLaughlin and
  Bader~\cite{McLaughlin:2014:SAH} uses task
  parallelism, dynamic load balancing, and sampling techniques to
  perform BC computation in parallel from different sources on different GPU
  SMXs.

\item[Gunrock Implementation] Gunrock's implementation also contains
  two phases. The first phase has an advance step identical to the
  original BFS and a computation step that computes the number of
  shortest paths from source to each vertex. The second phase uses an
  advance step to iterate over the BFS frontier backwards with a
  computation step to compute the dependency scores. We achieve
  competitive performance on scale-free graphs with the latest
  hardwired BC algorithm~\cite{McLaughlin:2015:AFE}. Within Gunrock,
  we haven't yet considered task parallelism since it appears to be
  limited to BC, but it is an interesting area for future work.
\end{description}

\subsection{Connected Component Labeling}

The connected component primitive labels the vertices in each
connected component in a graph with a unique component ID\@.

\begin{description}[style=unboxed, leftmargin=0cm]
\item[Hardwired GPU Implementation] Soman et al.~\cite{Soman:2010:AFG}
  base their implementation on two PRAM algorithms: hooking and
  pointer-jumping. Hooking takes an edge as the input and tries to set
  the component IDs of the two end vertices of that edge to the same
  value. In odd-numbered iterations, the lower vertex writes its value
  to the higher vertex, and vice versa in the even numbered iteration.
  This strategy increases the rate of convergence. Pointer-jumping
  reduces a multi-level tree in the graph to a one-level tree (star).
  By repeating these two operators until no component ID changes for
  any node in the graph, the algorithm will compute the number of
  connected components for the graph and the connected component to
  which each node belongs.

\item[Gunrock Implementation] Gunrock uses a filter operator on an edge
    frontier to implement hooking. The frontier starts with all edges and
    during each iteration, one end vertex of each edge in the frontier tries to
    assign its component ID to the other vertex, and the filter step removes
    the edge whose two end vertices have the same component ID\@.  We repeat
    hooking until no vertex's component ID changes and then proceed to
    pointer-jumping, where a filter operator on vertices assigns the component
    ID of each vertex to its parent's component ID until it reaches the root.
    Then a filter step removes the node whose component ID equals its own node
    ID\@. The pointer-jumping phase also ends when no vertex's component ID
    changes.
\end{description}

\subsection{PageRank and Other Node Ranking Algorithms}

The PageRank link analysis algorithm assigns a numerical weighting to
each element of a hyperlinked set of documents, such as the World Wide
Web, with the purpose of quantifying its relative importance within
the set. The iterative method of computing PageRank gives each vertex
an initial PageRank value and updates it based on the PageRank of its
neighbors, until the PageRank value for each vertex converges.
PageRank is one of the simplest graph algorithms to implement on GPUs
because the frontier always contains all vertices, so its computation
is congruent to sparse matrix-vector multiply; because it is simple,
most GPU frameworks implement it in a similar way and attain similar
performance.

In Gunrock, we begin with a frontier that contains all vertices in the
graph and end when all vertices have converged. Each iteration
contains one advance operator to compute the PageRank value on the
frontier of vertices, and one filter operator to remove the vertices
whose PageRanks have already converged. We accumulate PageRank values
with AtomicAdd operations.

\begin{description}[style=unboxed, leftmargin=0cm]
\item[Bipartite graphs] Geil et al.~\cite{Geil:2014:WGC:nourl} used
  Gunrock to implement Twitter's who-to-follow algorithm
  (``Money''~\cite{Goel:2014:TWT}), which incorporated three
  node-ranking algorithms based on bipartite graphs (Personalized
  PageRank, Stochastic Approach for Link-Structure Analysis (SALSA),
  and Hyperlink-Induced Topic Search (HITS))\@. Their implementation,
  the first to use a programmable framework for bipartite graphs,
  demonstrated that Gunrock's advance operator is flexible enough to
  encompass all three node-ranking algorithms, including a 2-hop
  traversal in a bipartite graph.
\end{description}

Beyond the five graph primitives we evaluate here, we have developed
or are actively developing several other graph primitives in Gunrock,
including minimal spanning tree, maximal independent set, graph
coloring, Louvain's method for community detection, and graph
matching.

\section{Experiments \& Results}
\label{sec:performance-and-analysis}

\begin{table}
  \small
  \centering
  \setlength{\tabcolsep}{3pt}
  \begin{tabular}{*{6}{c}} \toprule Dataset &Vertices&Edges&Max Degree& Diameter& Type \\
    \midrule
    soc-orkut & 3M & 212.7M & 27,466 & 9 & rs
    \\ hollywood-09 & 1.1M & 112.8M & 11,467 & 11 & rs
    \\ indochina-04 & 7.4M & 302M & 256,425 & 26 & rs
    \\ kron\_g500-logn21 & 2.1M & 182.1M & 213,904 & 6 & gs
    \\ rgg\_n\_24 & 16.8M & 265.1M & 40 & 2622 & gm
    \\ roadnet\_CA & 2M & 5.5M & 12 & 849 & rm
    \\ \bottomrule
  \end{tabular}
  \caption{Dataset Description Table. Graph types are: r: real-world, g: generated, s: scale-free, and m: mesh-like. \label{tab:dataset}}
\end{table}

We ran all experiments in this paper on a Linux workstation with
2$\times$3.50~GHz Intel 4-core, hyperthreaded E5-2637 v2 Xeon CPUs, 528~GB of
main memory, and an NVIDIA K40c GPU with 12~GB on-board memory.  GPU programs
were compiled with NVIDIA's nvcc compiler (version~7.0.27) with the -O3 flag.
The BGL and PowerGraph code were compiled using gcc 4.8.4 with the -O3 flag.
Ligra was compiled using icpc 15.0.1 with CilkPlus. All results ignore transfer
time (both disk-to-memory and CPU-to-GPU)\@. All tests were run 10 times with
the average runtime used for results.

The datasets used in our experiments are shown in Table~\ref{tab:dataset}. We
converted all datasets to undirected graphs. The six datasets include both
real-world and generated graphs; the topology of these datasets spans from
regular to scale-free.

Soc-orkut (soc) and hollywood-09 (h09) are two social graphs; indochina-04
(i04) is a crawled hyperlink graph from indochina web domains; kron\_g500-logn21
(kron) is a generated R-MAT graph. All four datasets are scale-free graphs with
diameters of less than 20 and unevenly distributed node degrees (80\% of nodes
have degree less than 64).

Both rgg\_n\_24 (rgg) and roadnet\_CA (roadnet) datasets have large diameters
with small and evenly distributed node degrees (most nodes have degree less than 12).

soc is from Network Repository; i04, h09, and kron are from UF Sparse Matrix Collection; rgg is a random geometric graph we generated.

The edge weight values (used in SSSP) for each dataset are random values
between 1 and 64.

\begin{table}
\small
    \centering
        \begin{tabular}{*{5}{c}} \toprule Algorithm & Galois &BGL&PowerGraph&Medusa \\
        \midrule
        BFS & 2.811 & --- & --- & 6.938
        \\ SSSP & 0.725 & 52.04 & 6.207 & \emph{11.88}
        \\ BC & 1.494 & --- & --- & ---
        \\ PageRank & 1.94 & 337.6 & 9.683 & \emph{8.982}
        \\ CC & 1.859 & 171.3 & 143.8 & ---
        \\ \bottomrule
    \end{tabular}
    \caption{Geometric-mean runtime speedups of Gunrock on the
      datasets from Table~\ref{tab:dataset} over frameworks
      not in Table~\ref{tab:exp_largetable}. Due to Medusa's memory
      limitations, its SSSP and PageRank comparisons were measured on
      smaller datasets.\label{tab:speedup}}
\end{table}

\begin{table*}[t]
  \small
  \centering
  \renewcommand{\arraystretch}{1.0} 
    \begin{tabular}{*{13}{c}} \toprule
      && \multicolumn{5}{c}{Runtime (ms) [lower is better]} && \multicolumn{5}{c}{Edge throughput (MTEPS) [higher is better]} \\
      \cmidrule{3-7}\cmidrule{9-13}
      &         &           &         & Hardwired &       &      &&     &                  & Hardwired &    &     \\
      Alg. & Dataset & CuSha & MapGraph & GPU       & Ligra & Gunrock && CuSha & MapGraph & GPU       & Ligra & Gunrock \\
      \midrule\parbox[t]{2mm}{\multirow{6}{*}{\rotatebox[origin=c]{90}{BFS}}} & soc & 251.8 & OOM & 45.43 & 27.2 & 47.23
      && 844.7 & --- & 4681 & 7819 & 4503
      \\ & h09 & 244.1 & 62.9 & 22.43 & 13.9 & 20.01
      && 461.5 & 1791 & 5116 & 8100 & 5627
      \\ & i04 & 1809 & OOM & 84.08 & 223 & 62.12
      && 164.8 & --- & 4681 & 1337 & 4799
      \\ & kron & 237.9 & 162.7 & 37.33 & 18.5 & 19.15
      && 765.2 & 1119 & 4877 & 9844 & 9510
      \\ & rgg & 52522 & OOM & 202.5 & 1020 & 351.4
      && 5.048 & --- & 1309 & 260 & 754.6
      \\ & roadnet  & 288.5 & 61.66 & 8.21 & 82.1 & 31
      && 19.14 & 89.54 & 672.9 & 67.25 & 178.1
      \\ \midrule\parbox[t]{2mm}{\multirow{6}{*}{\rotatebox[origin=c]{90}{SSSP}}} & soc & --- & OOM & 1106.6\textbf{*} & 950 & 1088
      && --- & --- & --- & --- & 195.5
      \\ & h09 & 1043 & OOM & 308.5\textbf{*} & 281 & 100.4
      && --- & --- & --- & --- & 1122
      \\ & i04 & --- & OOM & OOM & 850 & 511.5
      && --- & --- & --- &--- & 582.9
      \\ & kron & 315.5 & 540.8 & 677.7\textbf{*} & 416 & 222.7
      &&--- & ---& --- &--- & 817.6
      \\ & rgg & --- & OOM & OOM & 103000 & 117089
      &&--- & --- & --- &--- & 2.264
      \\ & roadnet & 1185 & 1285 & 224.2 & 451 & 222.1
      && ---& ---& 24.63 & --- &  24.86
      \\ \midrule\parbox[t]{2mm}{\multirow{6}{*}{\rotatebox[origin=c]{90}{BC}}} & soc & --- & --- & 1044 & 223 & 721.2
      && ---& ---& 407.4 & 1907 & 589.8
      \\ & h09 &--- & --- & 479.5 & 78.6 & 132.3
      && ---& ---& 469.6& 2867 & 1703
      \\ & i04 & --- & --- & 1389 & 557 & 164.3
      &&--- & ---& 429.1 & 1071 & 3630
      \\ & kron & --- & --- & 488.3 & 184 & 716.1
      &&--- & ---& 745.8 &1979 & 508.5
      \\ & rgg & --- & --- & 25307 & 2720 & 1449
      &&--- & ---& 20.94 &195 & 366
      \\ & roadnet & --- & --- & 256.8 & 232 & 120.6
      &&--- &--- & 42.99 &47.6 & 91.57
      \\ \midrule\parbox[t]{2mm}{\multirow{6}{*}{\rotatebox[origin=c]{90}{PageRank}}} & soc & 105.8 & OOM & --- &721 & 176
      \\ & h09 &43.27 & 94.35 & --- & 107 & 27.31
      \\ & i04 &121.8 & OOM & --- & 273 & 74.28
      \\ & kron &46.6 & 739.8 & --- & 456 & 176.2
      \\ & rgg & 48.6 & OOM & --- & 307 & 80.42
      \\ & roadnet &0.864 & 8.069 & --- & 14.6 & 6.691
      \\ \midrule\parbox[t]{2mm}{\multirow{6}{*}{\rotatebox[origin=c]{90}{CC}}} & soc & --- & --- & 91.58 & 313 & 252.9
      \\ & h09& --- & --- & 37.05 & 129 &  202.8
      \\ & i04  & --- & --- & 120.8 & 535 &  2501
      \\ & kron & --- & --- & 142.7 & 311 &  428.9
      \\ & rgg & --- & --- & 109.6 & 3280 &  552.7
      \\ & roadnet& --- & --- & 6.78 & 776 &  25.52
      \\ \bottomrule
    \end{tabular}
  \caption{Gunrock's performance comparison (runtime and edge throughput) with other graph libraries (CuSha, MapGraph, Ligra) and hardwired GPU implementations.
 SSSP MTEPS statistics are unavailable in most frameworks. All PageRank times are normalized to one iteration. Hardwired GPU implementations for each primitive are b40c (BFS)~\protect\cite{Merrill:2012:SGG}, delta-stepping SSSP~\protect\cite{Davidson:2014:WPG:nourl} (numbers with \textbf{*} are achieved without delta-stepping optimization, otherwise will run out of memory), gpu\_BC (BC)~\protect\cite{Sariyuce:2013:BCO}, and conn (CC)~\protect\cite{Soman:2010:AFG}. OOM means out-of-memory.\label{tab:exp_largetable}}
\end{table*}

\begin{figure*}[ht]
    \centering
    \includegraphics[width=0.9\textwidth]{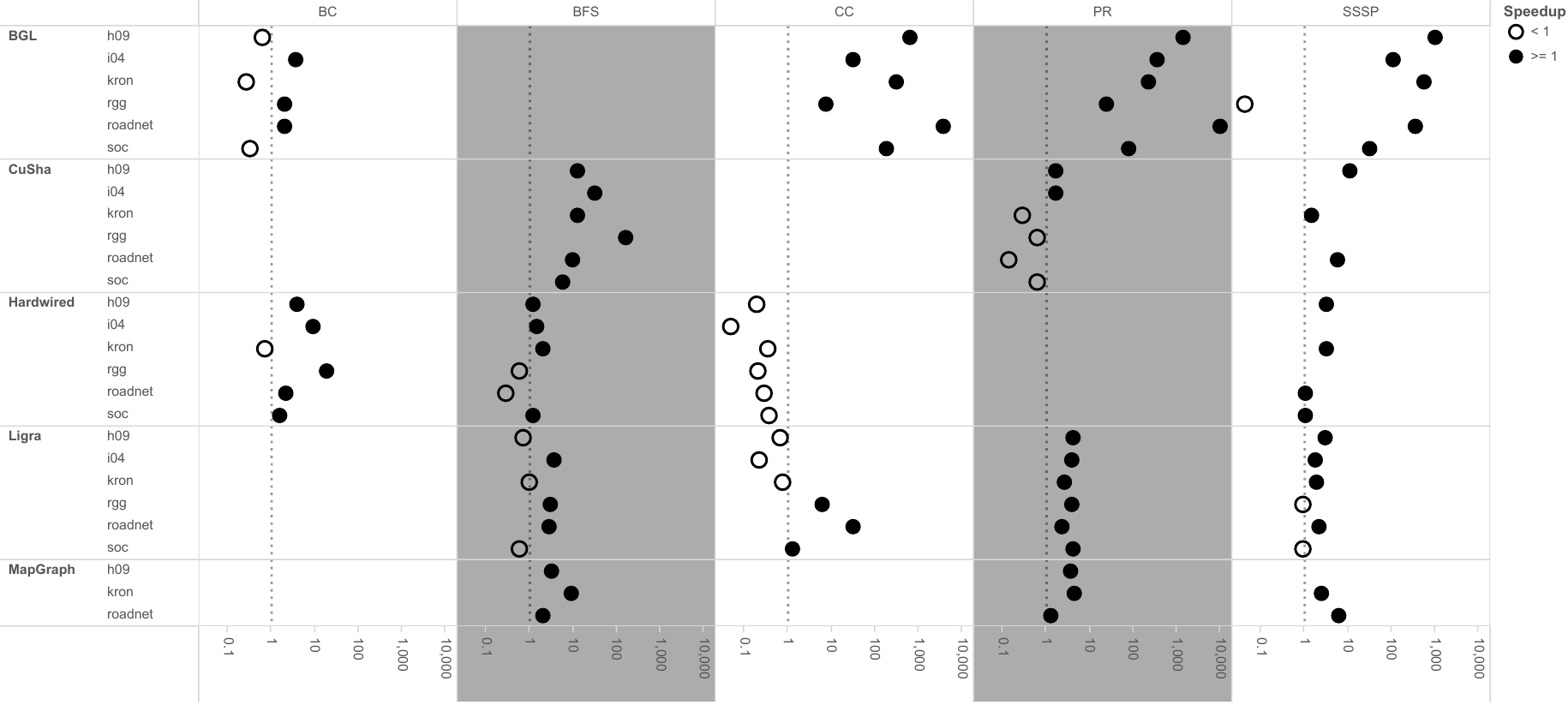}
    \centering
    \caption{Execution-time speedup for Gunrock vs.\ five other graph
      processing libraries/hardwired algorithms on six different
      graph inputs. Data is from Table~\ref{tab:exp_largetable}.
      Black dots indicate Gunrock is faster, white dots
      slower.\label{fig:speedup}}
\end{figure*}

\begin{description}[style=unboxed, leftmargin=0cm]
\item[Performance Summary] Tables~\ref{tab:speedup} and~\ref{tab:exp_largetable}, and Figure~\ref{fig:speedup}, compare
Gunrock's performance against several other graph libraries and hardwired GPU implementations. In
  general, Gunrock's performance on BFS-based primitives (BFS, BC, and SSSP)
  shows comparatively better results when compared to other graph libraries on
  four scale-free graphs (soc, h09, i04, and kron), than on two small-degree
  large-diameter graphs, rgg and roadnet. The primary reason is our
  load-balancing strategy during traversal (Table~\ref{tab:wee} shows Gunrock's
  superior performance on warp efficiency, a measure of load-balancing quality,
  across GPU frameworks and datasets), and particularly our emphasis on good
  performance for highly irregular graphs. As well, graphs with uniformly low
  degree expose less parallelism and would tend to show smaller gains in
  comparison to CPU-based methods.

\item[vs.\ CPU Graph Libraries] We compare Gunrock's performance with four CPU
    graph libraries: the Boost Graph Library (BGL)~\cite{Siek:2001:TBG}, one of
    the highest-performing CPU single-threaded graph
    libraries~\cite{McColl:2014:APE}; PowerGraph, a popular distributed graph
    library~\cite{Gonzalez:2012:PDG}; and Ligra~\cite{Shun:2013:LAL} and
    Galois~\cite{Nguyen:2013:ALI,Pingali:2011:TTO}, two of the
    highest-performing multi-core shared-memory graph libraries. Against both
    BGL and PowerGraph, Gunrock achieves 6x--337x speedup on average on all
    primitives. Compared to Ligra, Gunrock's performance is generally
    comparable on most tested graph primitives; note Ligra uses both CPUs
    effectively. The performance inconsistency for SSSP vs.\ Ligra is due to
    comparing our Dijkstra-based method with Ligra's Bellman-Ford algorithm.
    Our SSSP's edge throughput is smaller than BFS but similar to BC because of
    similar computations (atomicMin vs.\ atomicAdd) and a larger number of
    iterations for convergence. The performance inconsistency for BC vs.\ Ligra
    on four scale-free graphs is because that Ligra applies pull-based
    traversal on BC while Gunrock has not yet done so. Compared to Galois,
    Gunrock shows more speedup on traversal-based graph primitives (BFS, SSSP,
    and BC) and less performance advantage on PageRank and CC due to their
    dense computation and more regular frontier structures.

\item[vs.\ Hardwired GPU Implementations and GPU Libraries] Compared to
    hardwired GPU implementations, depending on the dataset, Gunrock's
    performance is comparable or better on BFS, BC, and SSSP\@. For CC, Gunrock
    is 5x slower (geometric mean) than the hardwired GPU implementation due to
    irregular control flow because each iteration starts with full edge lists
    in both hooking and pointer-jumping phases. The alternative is extra steps
    to perform additional data reorganization. This tradeoff is not typical of
    our other primitives. While still achieving high performance, Gunrock's
    application code is smaller in size and clearer in logic compared to other
    GPU graph libraries\footnote{We believe this assertion is true given our
        experience with other GPU libraries when preparing this evaluation
    section, but freely acknowledge this is nearly impossible to quantify. We
invite readers to peruse our annotated code for BFS and SALSA at
\url{http://gunrock.github.io/gunrock/doc/annotated_primitives/annotated_primitives.html}.}.
Gunrock's Problem class (that defines problem data used for the graph
algorithm) and kernel enactor are both template-based C++ code; Gunrock's
functor code that specifies per-node or per-edge computation is C-like device
code without any CUDA-specific keywords. For a new graph primitive, users only
need to write from 133 (simple primitive, BFS) to 261 (complex primitive,
SALSA) lines of code. Writing Gunrock code may require parallel programming
concepts (e.g., atomics) but neither details of low-level GPU programming nor
optimization knowledge.

Gunrock compares favorably to existing GPU graph libraries. MapGraph is faster
than Medusa on all but one test~\cite{Fu:2014:MAH} and Gunrock is faster than
MapGraph on all tests: the geometric mean of Gunrock's speedups over MapGraph
on BFS, SSSP, and PageRank are
4.3, 3.7, and 2.1, respectively. Gunrock also outperforms CuSha on BFS and
  SSSP\@. For PageRank, Gunrock achieves comparable performance without the
  G-Shard data preprocessing, which serves as the main load-balancing module in
  CuSha. The 1-GPU Gunrock implementation has 1.83x more MTEPS (4731 vs. 2590)
  on direction-optimized BFS on the soc-LiveJournal dataset (a smaller
  scale-free graph in their test set) than the 2-CPU, 2-GPU configuration of
  Totem~\cite{Sallinen:2015:ADB}.  All three GPU BFS-based
  high-level-programming-model efforts (Medusa, MapGraph, and Gunrock) adopt
  load-balancing strategies from Merrill et al.'s BFS~\cite{Merrill:2012:SGG}.
  While we would thus expect Gunrock to show similar performance on BFS-based
  graph primitives as these other frameworks, we attribute our performance
  advantage to two reasons: (1) our improvements to efficient and load-balanced
  traversal that are integrated into the Gunrock core, and (2) a more powerful,
  GPU-specific programming model that allows more efficient high-level graph
  implementations. (1) is also the reason that Gunrock implementations can
  compete with hardwired implementations; we believe Gunrock's load-balancing
  and work distribution strategies are at least as good as if not better than
  the hardwired primitives we compare against. Gunrock's memory footprint is at
  the same level as Medusa and better than MapGraph (note the OOM test cases
  for MapGraph in Table~\ref{tab:exp_largetable}). The data size is
  $\alpha|E|+\beta|V|$ for current graph primitives, where $|E|$ is the number of
  edges, $|V|$ is the number of nodes, and $\alpha$ and $\beta$ are both integers
  where $\alpha$ is usually
  1 and at most 3 (for BC) and $\beta$ is between 2 to 8.
\end{description}

\begin{figure*}
    \centering
    \includegraphics[width=0.27\textwidth]{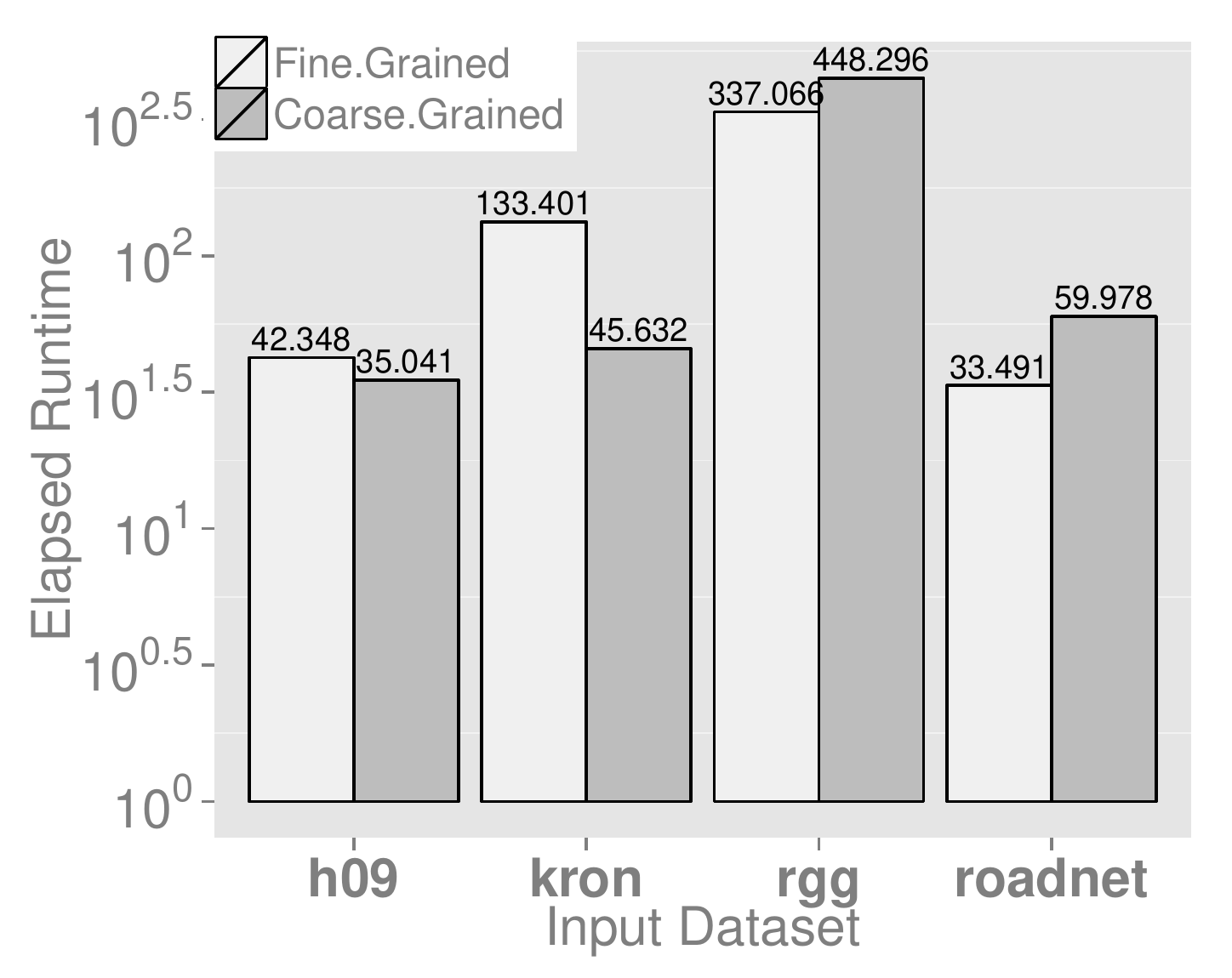}
    \includegraphics[width=0.27\textwidth]{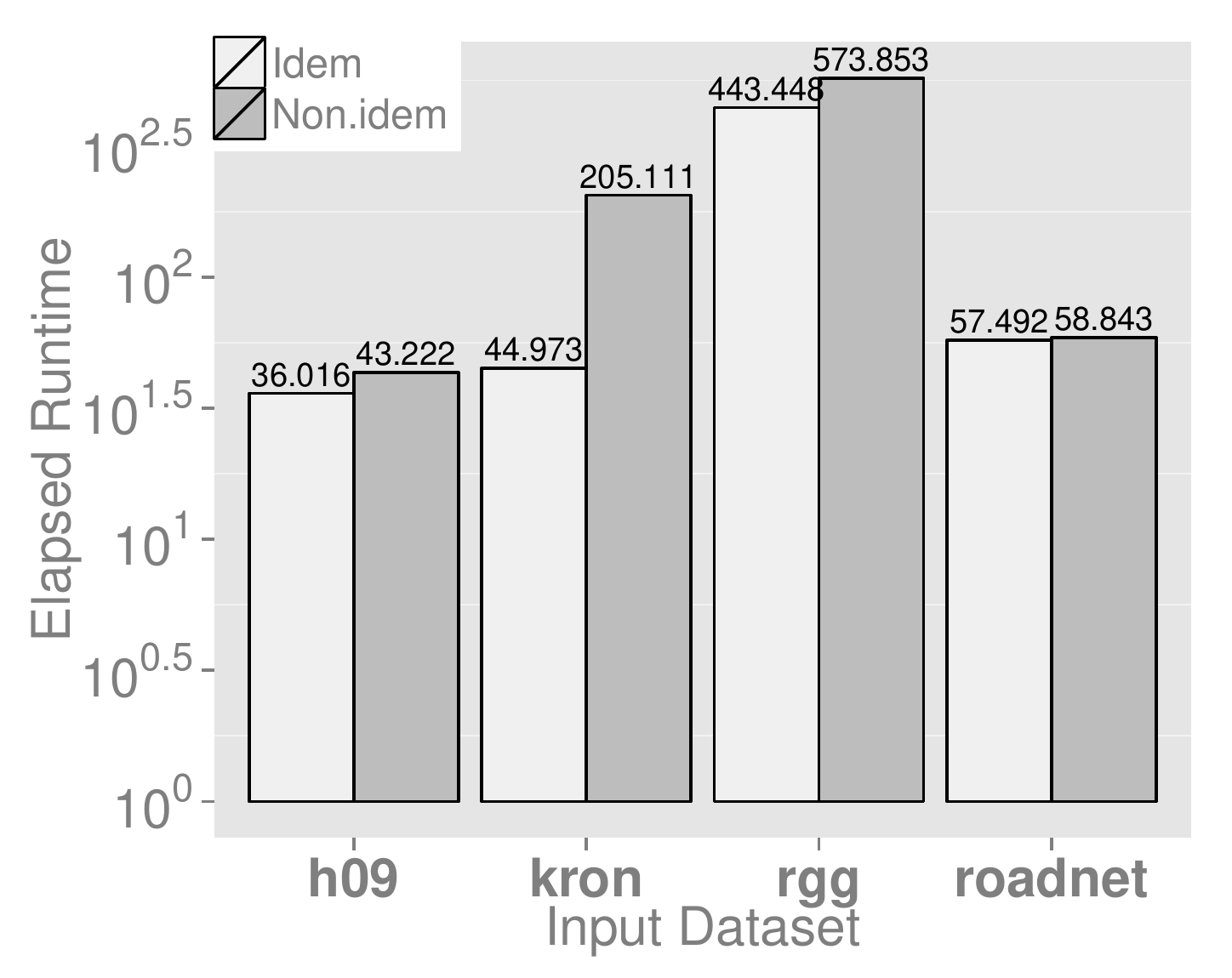}
    \includegraphics[width=0.27\textwidth]{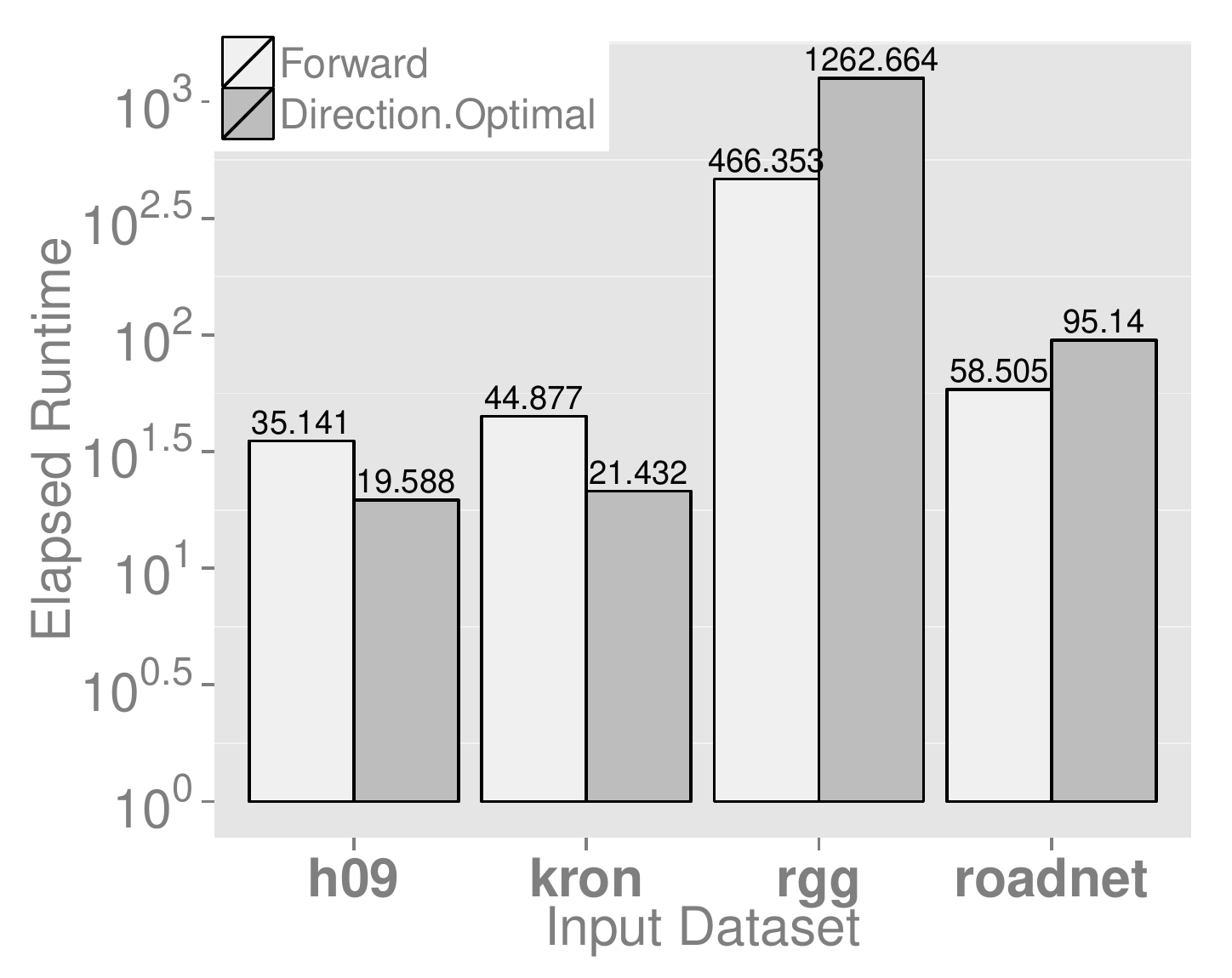}
    \centering \caption{Left: Performance comparison with two
      different workload mapping optimizations. Middle: Performance
      comparison on graph traversal with idempotent operations enabled
      vs.\ disabled. Right: Performance comparison between forward and
      direction optimal graph traversal.\label{fig:exp_optimization}}
\end{figure*}

Figure~\ref{fig:exp_optimization} shows how different optimization strategies
improve the performance of graph traversal; here we use BFS as an example. As noted in
Section~\ref{sec:frame:optimize}, the load-balancing traversal method works
better on social graphs with irregular distributed degrees, while the
Thread-Warp-CTA method works better on graphs where most nodes have small
degrees. The direction-optimal traversal strategy also works better on social
graphs, whereas on the road-network and bitcoin-transactions graph, we see less
concurrent discovery and the performance benefits are not as significant. In
general, we can predict which strategies will be most beneficial based only on
the degree distribution; many application scenarios may allow precomputation of
this distribution and thus we can choose the optimal strategies before we begin
computation.

\label{sec:warp-execution-efficiency}
\begin{table}
  \centering
  \resizebox{\linewidth}{!}{
  \begin{tabular}{@{}llcccccc@{}}
    \toprule
    Alg. & Framework & soc & h09  & i04       & kron      & rgg & roadnet \\
    \midrule
    \multirow{3}{*}{BFS}
    & Gunrock & 97.39\% & 97.35\% & 97.97\% & 97.73\% & 96.72\% & 97.01\% \\
    & MapGraph & --- & 95.81\% & --- & 97.19\% & --- & 87.49\% \\
    & CuSha & 77.12\% & 80.12\% & 72.40\% & 50.34\%     & 85.32\% & 87.80\% \\
    \midrule
    \multirow{3}{*}{SSSP}
    & Gunrock & 83.35\% & 82.56\% & 83.18\% & 85.15\% & 82.84\% & 83.47\% \\
    & MapGraph & --- & --- & --- & 95.62\% & --- & 91.51\% \\
    & CuSha & 78.40\% & 80.17\% & 76.63\% & 52.72\% & 86.96\% & 85.28\% \\
    \midrule
    \multirow{3}{*}{PR}
    & Gunrock & 99.56\% & 99.42\% & 99.54\% & 99.43\% & 99.52\% & 99.49\% \\
    & MapGraph & --- & 98.97\% & --- & 99.16\% & --- & 96.27\% \\
    & CuSha & 82.29\% & 87.26\% & 85.10\% & 63.46\% & 91.04\% & 89.23\% \\
    \bottomrule
  \end{tabular}}
\caption{Average warp execution efficiency (fraction of threads
  active during computation). This figure is a good metric for
  the quality of a framework's load-balancing capability. (--- indicates
  the graph framework ran out of memory.) \label{tab:wee}}
\end{table}

\section{Future Work}
\label{sec:futurework}
We believe Gunrock currently provides an excellent foundation for
developing GPU-based graph primitives. We hope to extend Gunrock
with the following improvements:

\begin{description}[style=unboxed, leftmargin=0cm]

\item[Dynamic graphs] While Gunrock currently implements several graph
  primitives (e.g., minimum spanning tree and connected components)
  that internally modify graph topology, generalized support of
  dynamic graphs on GPUs that change their structure during
  computation is still an unsolved problem.

\item[Global, neighborhood, and sampling operations] In Gunrock,
  computation on vertices and edges is convenient and fast. However,
  global and neighborhood operations, such as reductions over neighbor
  lists, generally require less-efficient atomic operations and are
  an ongoing challenge. We believe a new gather-reduce operator on
  neighborhoods associated with vertices in the current frontier both
  fits nicely into Gunrock's abstraction and will significantly
  improve performance on this operation. We also expect to explore a
  ``sample'' step that can take a random subsample of a frontier,
  which we can use to compute a rough or seeded solution that may
  allow faster convergence on a full graph.

\item[Kernel fusion] Gunrock's implementation generally allows more
  opportunities to fuse multiple operations into a single kernel than
  GAS+GPU implementations (Section~\ref{sec:gunrock-kernel-fusion}),
  but does not achieve the level of fusion of hardwired
  implementations. This interesting (and unsolved, in the general
  case) research problem represents the largest performance gap
  between hardwired and Gunrock primitives.

\item[Scalability] Today, Gunrock is implemented for single-GPU computation
  and graphs that fit into the GPU's memory. We believe the
  contributions in this paper successfully advance the state of the
  art on one GPU, but for greater impact, a future Gunrock must scale
  in three directions: to leverage the larger memory capacity of a CPU
  host; to multiple GPUs on a single node; and to a distributed,
  multi-node clustered system. Current GPU work in this area generally
  targets only specific primitives (e.g., Merrill et al.'s. multi-GPU
  BFS~\cite{Merrill:2012:SGG}) and/or is not performance-competitive
  with large-memory, single-node CPU implementations. We hope that
  Gunrock's data-centric focus on frontiers---which we believe is
  vital for data distributions that go beyond a single GPU's
  memory---provides an excellent substrate for a future scalable
  GPU-based graph implementation.

\end{description}

\section{Conclusions}
\label{sec:conclusion}
Gunrock was born when we spent two months writing a single hardwired
GPU graph primitive. We knew that for GPUs to make an impact in graph
analytics, we had to raise the level of abstraction in building graph
primitives. From the beginning, we designed Gunrock with the GPU in
mind, and its data-centric, frontier-focused abstraction has proven to map
naturally to the GPU, giving us both good performance and good
flexibility. We have also found that implementing this abstraction has
allowed us to integrate numerous optimization strategies, including
multiple load-balancing strategies for traversal, direction-optimal
traversal, and a two-level priority queue. The result is a framework
that is general (able to implement numerous simple and complex graph
primitives), straightforward to program (new primitives only take a
few hundred lines of code and require minimal GPU programming
knowledge), and fast (on par with hardwired primitives and faster than
any other programmable GPU graph library).

\section*{Acknowledgments}
\label{acks}
We thank Joe Mako for providing the speedup chart design. Also, thanks to the
DARPA XDATA program and our program managers Christopher White and Wade Shen
(US Army award W911QX-12-C-0059); DARPA STTR awards D14PC00023 and
D15PC00010\@; NSF awards CCF-1017399 and OCI-1032859; and UC Lab Fees Research
Program Award 12-LR-238449\@.


\let\doi\relax                  
\bibliographystyle{abbrv}
\bibliography{bib/all,gunrock}

\clearpage
\appendix
\section{Artifact description}

\subsection{Abstract}

{\em The artifact contains all the executables of the current existing graph primitives in Gunrock's latest version on github, as well as the shell scripts of running them. It can support the runtime and/or edge throughput results in Table 3 of our PPoPP'2016 paper \textbf{Gunrock: A High-Performance Graph Processing Library on the GPU}. To validate the results, run the test scripts and check the results piped in the according text output files.}

\subsection{Description}

\subsubsection{Check-list (artifact meta information)}

{\small
\begin{itemize}
  \item {\bf Algorithm: breadth-first search, single-source shortest path, betweenness centrality, Pagerank, connected component}
  \item {\bf Program: CUDA and C/C++ code}
  \item {\bf Compilation: Host code: gcc 4.8.4 with the -O3 flag; device code: nvcc 7.0.27 with the -O3 flag}
  \item {\bf Binary: CUDA executables}
  \item {\bf Data set: Publicly available matrix market files}
  \item {\bf Run-time environment: Ubuntu 12.04 with CUDA and GPU Computing SDK installed}
  \item {\bf Hardware: Any GPU with compute capability $\geq$ 3.0 (Recommended GPU: NVIDIA K40c GPU)}
  \item {\bf Output: Runtime and/or edge throughput}
  \item {\bf Experiment workflow: Git clone project; download the datasets; run the test scripts; observe the results}
  \item {\bf Publicly available?: Yes}
\end{itemize}
}

\subsubsection{How delivered}

{\em Gunrock is an open source library under Apache 2.0 license and is hosted with code, API specifications, build instructions, and design documentations on Github.}

\subsubsection{Hardware dependencies}
{\em Gunrock requires NVIDIA GPU with the compute capability of no less than 3.0.}

\subsubsection{Software dependencies}
{\em Gunrock requires Boost (for CPU reference) and CUDA with version no less than 5.5. Gunrock has been tested on Ubuntu 12.04/14.04, and is expected to run correctly under other Linux distributions.}

\subsubsection{Datasets}
All datasets are either publicly available or generated using standard graph generation software. Users will be able to run script to get these datasets once they built Gunrock code. The rgg graph is generated by Gunrock team. The download link
is provided here: \url{https://drive.google.com/uc?export=download&id=0Bw6LwCuER0a3VWNrVUV6eTZyeFU}. Please located the unzipped \url{rgg_n_2_24_s0.mtx} file under \url{gunrock_dir/datasets/large/rgg_n_2_24_s0/}. Users are welcom to try other datasets or generate rgg/R-MAT graphs using the command line option during the test. We currently only support matrix market format files as input. 

\subsection{Installation}

{\em Follow the build instruction on Gunrock's github page (\url{http://gunrock.github.io/}), users can build Gunrock and generate the necessary executables for the experiments.}

\subsection{Experiment workflow}

{\em For the convenience of the artifact evaluation, we provide a series of shell scripts which run the graph primitives we have described in the paper and store the results in the output text files. Below are the steps to
download Gunrock code, build, run the experiments, and observe the results.}
\begin{itemize}
\item[-]
Clone Gunrock code to the local machine:
\begin{lstlisting}[language=bash]
$ git clone https://github.com/gunrock/gunrock.git
$ cd gunrock
$ git submodule init && git submodule update
\end{lstlisting}

\item[-]
Use CMake to build Gunrock. Make sure that boost and CUDA is correctly installed before this step:
\begin{lstlisting}[language=bash]
$ cd /path/to/gunrock/../
$ mkdir gunrock_build && cd gunrock_build
$ cmake ../gunrock/
$ make -j16
\end{lstlisting}
The last comand will build Gunrock's executables under \url{gunrock_build/bin} and shared library under \url{gunrock_build/lib}.

\item[-] Prepare the dataset. First step into Gunrock directory:
\begin{lstlisting}[language=bash]
$ cd /path/to/gunrock/
$ cd dataset/large/ && make
\end{lstlisting}
This will download and extract all the large datasets, including the 6 datasets in the paper.

\item[-] Step into the test script directory and run scripts for five graph primitives:
\begin{lstlisting}[language=bash]
$ cd ../test-scripts
$ sh ppopp16-test.sh
\end{lstlisting}

\item[-] Observe the results for each dataset under five directories: BFS, SSSP, BC, PR, and CC.
\end{itemize}

\subsection{Evaluation and expected result}

{\em For BFS and SSSP, the expected results include both runtime and edge throughput. For BC, Pagerank, and CC, the expected results contain runtime only.}

\subsection{Notes}
{\em To know more about our library, send feedback, or file issues, please visit our github page (\url{http://gunrock.github.io/}).}

\end{document}